\documentclass[12pt]{iopart}   
\usepackage{graphicx}
\usepackage{iopams}    
\usepackage{setstack}  
\usepackage{cite}
\begin{document}
%
%
\title[A general scaling relation for Nb$_3$Sn]{A general scaling relation for the critical current density in Nb$_3$Sn}
\author{A Godeke$^{1,2}$\footnote{Now at Ernest Orlando Lawrence Berkeley National Laboratory, Berkeley, CA 94720 \\ Electronic address: \texttt{agodeke@lbl.gov}}, B ten Haken$^1$, H H J ten Kate$^1$ and D C Larbalestier$^2$}
\address{$^1$Low Temperature Division, Faculty of Science and  Technology, University of Twente, P.O. Box 217, 7500AE Enschede, The Netherlands}
\address{$^2$Applied Superconductivity Center, University of  Wisconsin, 1500 Engineering Drive, Madison, Wisconsin 53706, USA}
%
%
%
\begin{abstract}
We review the scaling relations for the critical current density ($J_{\rm c}$) in Nb$_3$Sn wires and include recent findings on the variation of the upper critical field ($H_{\rm c2}$) with temperature ($T$) and A15 composition. Measurements of $H_{\rm c2}(T)$ in inevitably inhomogeneous wires, as well as analysis of literature results, have shown that all available $H_{\rm c2}(T)$ data can be accurately described by a single relation from the microscopic theory. This relation also holds for inhomogeneity averaged, effective, $H_{\rm c2}^*(T)$ results and can be approximated by $H_{\rm c2}(t)/H_{\rm c2}(0) \cong 1-t^{1.52}$, with $t=T/T_{\rm c}$. Knowing $H_{\rm c2}^*(T)$ implies that also $J_{\rm c}(T)$ is known. We highlight deficiencies in the Summers/Ekin relations, which are not able to account for the correct $J_{\rm c}(T)$ dependence. Available $J_{\rm c}(H)$ results indicate that the magnetic field dependence for all wires from $\mu_0H \cong 1 \ \rm{T}$ up to about 80\% of the maximum $H_{\rm c2}$ can be described with Kramer's flux shear model, if non-linearities in Kramer plots when approaching the maximum $H_{\rm c2}$ are attributed to A15 inhomogeneities. The strain ($\epsilon$) dependence is introduced through a temperature and strain dependent $H_{\rm c2}^*(T,\epsilon)$ and Ginzburg-Landau parameter $\kappa_1(T,\epsilon)$ and a strain dependent critical temperature $T_{\rm c}(\epsilon)$. This is more consistent than the usual Ekin unification of strain and temperature dependence, which uses two separate and different dependencies on $H_{\rm c2}^*(T)$ and $H_{\rm c2}^*(\epsilon)$. Using a correct temperature dependence and accounting for the A15 inhomogeneities leads to the remarkable simple relation $J_{\rm c}(H,T,\epsilon)\cong(C/\mu_0H)s(\epsilon)(1-t^{1.52})(1-t^2)h^{0.5}(1-h)^2$, where $C$ is a constant, $s(\epsilon)$ represents the normalized strain dependence of $H_{\rm c2}^*(0)$ and $h=H/H_{\rm c2}^*(T,\epsilon)$. Finally, a new relation for $s(\epsilon)$ is proposed, which is an asymmetric version of our earlier deviatoric strain model and based on the first, second and third strain invariants. The new scaling relation solves a number of much debated issues with respect to $J_{\rm c}$ scaling in Nb$_3$Sn and is therefore of importance to the applied community, who use scaling relations to analyze magnet performance from wire results.
\end{abstract}
\submitto{\SUST Pre-print of Accepted Topical Review\\http://www.iop.org/EJ/journal/SUST}   
\maketitle
%
%
\section{\label{Introduction}Introduction}
Critical current measurements of Nb$_3$Sn wires are commonly interpolated and extrapolated using a set of empirical relations for $J_{\rm c}(H,T,\epsilon)$, generally referred to as Summers scaling~\cite{Summers1991}. Although these relations show acceptable accuracy for the description of limited data ranges~\cite{Godeke2002a}, large discrepancies appear between the fitted values for $T_{\rm c}(0)$ and $H_{\rm c2}(0)$ and actual measurements~\cite{Godeke2003}. Also at temperatures above about 13 K and for large compressive and tensile axial strains the Summers scaling is inaccurate. It is thus not suitable for precise extrapolations of measured results across a wide range. These deficiencies are generally recognized and lead to continuing discussions in the literature on improvements~\cite{Godeke2002a,Godeke2003,tenhaken1999,Cheggour2002,Keys2003,Cheggour1999,Godeke2002b,Godeke1999,Godeke2001,Oh2005a,Oh2005b}. Presently, much of the interpretation of measured results and ensuing discussions remain largely empirical and the proposed relations depend, for a significant part, on the initial assumptions. A notable exception is the recent work by Oh and Kim~\cite{Oh2005a,Oh2005b} who use Eliashberg theory~\cite{Eliashberg1960,Scalapino1966} in their attempt to arrive at a better founded scaling formalism.

As an alternative to empirical approaches, a series of relations resulting from microscopic theory exist for Type II superconductivity. Although the critical current cannot be related quantitatively to the microstructure, various bulk pinning force models have been developed~\cite{Matsushita1985,Kramer1973,Dew-Hughes1974,Dew-Hughes1987,Labusch1969a,Labusch1969b,Brandt1977,Campbell1972,Larkin1979}. These can be combined with theory on strain dependence~\cite{Markiewicz2004a,Markiewicz2004b,Markiewicz2004c} and the field-temperature phase boundary~\cite{deGennes1966,Parks1969,Maki1964,deGennes1964,Helfand1964,Helfand1966,Werthamer1966,Hohenberg1967,Werthamer1967,Eilenberger1967,Usadel1970,Usadel1971,Bergmann1973,Rainer1973,Rainer1974,Schopohl1981,Schopohl1985,Schossmann1986,Rieck1991,Arai2004,Kita2005}. The latter have been shown to reasonably describe measurements on well defined, quasi-homogeneous laboratory samples~\cite{Werthamer1967,Schopohl1985,Schossmann1986,Rieck1991,Arai2004,Orlando1979,Orlando1981,Foner1976,Foner1981a}.

It is questionable, however, whether such relations can directly describe the behavior of Nb$_3$Sn in technical wires due to the inevitable presence of inhomogeneities. Compositional inhomogeneities result from the A15 stability range from about 18 to 25~at.\%~Sn~\cite{Charlesworth1970}, combined with the solid state diffusion processes during the A15 formation reaction in wires. The resulting Sn gradients have been unambiguously detected in wires by compositional analysis~\cite{Wu1983,Hawes2000,Hawes2002,Godeke2005a,Abacherli2005}, property gradient measurements~\cite{Hawes2000,Hawes2002,Godeke2005a,Fischer2002a,Fischer2002b,Godeke2005b} and simulations~\cite{Godeke2005a,Godeke2005b,Marken1986a,Cooley2004}. These compositional inhomogeneities lead to significant changes in the electron-phonon interaction strength $\lambda_{\rm{ep}}$~\cite{Moore1979}. The interaction strength changes from the weak coupling BCS limit ($\lambda_{\rm{ep}}\ll1$) for 18 at.\% Sn to $\lambda_{\rm{ep}}\cong1.8$ for the stoichiometric composition~\cite{Orlando1979,Moore1979}. As a result, significant changes occur in $H_{\rm c2}(T)$ with Sn concentration~\cite{Orlando1979,Orlando1981,Moore1979,Godeke2006,Devantay1981,Flukiger1981,Jewell2004}. Since all stable A15 compositions from 18 to 25 at.\% Sn are observed in wires, a distribution of properties of about $5~\mathrm{T} \leq \mu_0 H_{\rm{c2}}(0) \leq 31~\mathrm{T}$ and $6~\mathrm{K} \leq T_{\mathrm{c}}(0) \leq 18~\mathrm{K}$ will occur over the A15 volumes~\cite{Godeke2005a}. In addition, it can be expected that also the strain state is far from homogeneous over the A15 volumes in a wire. These arguments render the validity of a full microscopic approach doubtful. Moreover, microscopic based relations can be relatively complex and use a significant number of parameters to describe the material involved. The main goal of this article is to find an improved scaling relation which balances the need for better founded formalisms and practical applicability to measured results in wires. As a starting point we use three central statements, which will be detailed further throughout the article.

Our first statement is that the temperature dependencies $H_{\rm c2}(T)$ and $H_{\rm c}(T)$ (the thermodynamic critical field) are known. We showed earlier~\cite{Godeke2005a,Godeke2005b} that the shape of $H_{\rm c2}(T)$ is constant for all available Nb$_3$Sn results, independent of the sample layout, morphology, A15 composition, strain state or applied critical state criterion and also holds for inhomogeneity averaged, extrapolated, $H_{\rm c2}^*(T)$ data. This means that the normalized temperature dependence of $H_{\rm c2}$ can be described by a single, known function from the microscopic theory. Also $H_{\rm c}(T)$ is known~\cite{Oh2005a,Tinkham1996,Guritanu2004} and $\kappa_1(T)$ immediately follows through $\kappa_1=H_{\rm c2}/(\surd2H_{\rm c})$~\cite{Maki1964}. This approach differs from the temperature dependence in the Summers relation, in which $H_{\rm c2}(T)$ is calculated from an estimated $\kappa_1(T)$~\cite{Summers1991}.

Our second statement is that non-linearities, which generally appear in Kramer plots~\cite{Kramer1973} when approaching the maximum $H_{\rm c2}$ that is present in a wire, can be attributed to A15 inhomogeneities. This assumption can be validated for specific wire layouts through a combination of simulations and measurements~\cite{Godeke2005b,Cooley2004}. Alternatively, one can chose to fit measured non-linearities assuming a different pinning model as is often done~\cite{Matsushita1985,Cheggour2002,Cheggour1999,Kroeger1980}. This will improve the quality of the fit per specific wire type when approaching the maximum $H_{\rm c2}$, but this is also the regime where $J_{\rm c}$ becomes very low. This regime is therefore of limited interest for applications. Fitting this regime per wire type will obviously result in a reduced overall error but also in a model with significantly reduced generality. We therefore chose not to include such non-linearities in the model fit to arrive at a generally valid relation for all Nb$_3$Sn wires, with a validity up to the regime where the effects of A15 inhomogeneity start to appear (typically from about 80\% of the maximum $H_{\rm c2}$).

Our third statement is that the introduction of strain sensitivity occurs solely through strain induced changes in $H_{\rm c2}$, $T_{\rm c}$ and $\kappa_1$. Alternative approaches can lead to a strain dependent pre-constant in the $J_{\rm c}(H,T,\epsilon)$ relation~\cite{Cheggour2002,Keys2003,Cheggour1999,Kroeger1980}, which implies that the pinning efficiency changes with strain. Although this cannot be ruled out, it can be argued that the required change in pinning efficiency of at least a factor of 2 over the relevant strain regime (below 1\% axial strain)~\cite{Cheggour1999,Kroeger1980}, is unrealistically large. A more reasonable approach to our opinion, is to assume that the main effect of strain is a modification of the electron-phonon interaction spectrum, which is supported by recent calculations~\cite{Oh2005a,Oh2005b,Markiewicz2004a,Markiewicz2004b,Markiewicz2004c}. This will obviously change directly $T_{\rm c}$, $H_{\rm c2}$ and $H_{\rm c}$ and thus $\kappa_1$. We will show below how this assumption, combined with a known temperature dependence of $H_{\rm c2}$ and $H_{\rm c}$ leads to a consistent introduction of strain and temperature effects in the $J_{\rm c}(H,T,\epsilon)$ relation by stating that $H_{\rm c2}^* \rightarrow H_{\rm c2}^*(T,\epsilon)$ and $\kappa_1 \rightarrow \kappa_1(T,\epsilon)$. Note that this is contrary to the usual unification of strain and temperature dependence, which assumes two separate and different dependencies on $H_{\rm c2}^*(T)$ and $H_{\rm c2}^*(\epsilon)$~\cite{Ekin1980}.

We will show below how these three starting points lead to a consistent, simple and accurate relation for $J_{\rm c}(H,T,\epsilon)$. In Section \ref{Descriptions}, we will summarize the descriptions that are available throughout the literature and arrive at a general form for $J_{\rm c}(H,T,\epsilon)$. In Section \ref{Comparison}, we will systematically compare this general form to measurements to determine the resulting scaling relation. The overall accuracy will be demonstrated in Section \ref{Overall} and discussed in Section \ref{Discussion}, together with a determination of the minimal required set of measurements that are needed to fully characterize a specific wire. Our conclusions are presented in Section \ref{Conclusions}.
%
%
\section{\label{Descriptions}The critical surface}
\subsection{\label{F-H} Magnetic field dependence}

\subsubsection{Relations for the Pinning Force}
The magnetic field dependence of the critical current density in Nb$_3$Sn is determined by de-pinning of the flux-line lattice and thus by the magnetic field dependence of the bulk pinning force $F_{\rm p} = -J_{\rm c} \times B$. The latter depends on the microscopic de-pinning mechanism. A general description for $F_{\rm p}(H)$ results from the observation that, for a multitude of low temperature superconductors, the bulk pinning force scales according to:
\begin{equation}
 \label{fietzwebb}
 F_{\rm p} \propto \frac{{H_{\rm c2}^\nu  }}{{\kappa _1^\gamma  }}f\left( h \right),
\end{equation}
as was first pointed out by Fietz and Webb~\cite{Fietz1969}. In \eref{fietzwebb}, $h$ is the reduced magnetic field $H/H_{\rm c2}$. The value for $\nu$ is often obtained by neglecting $\kappa_1$ and plotting the maximum bulk pinning force against $H_{\rm c2}$ on a double logarithmic scale. Values for $\nu$ found in this way are usually between 2 and 3 and for Nb$_3$Sn values of 2 and 2.5 are mostly cited. However, since $\kappa_1$ may change up to 50\% with temperature (Section \ref{microkappa}), its temperature dependence cannot be neglected for an accurate description of the critical surface of Nb$_3$Sn.

Several models for $f(h)$ exist~\cite{Kramer1973,Dew-Hughes1974,Dew-Hughes1987}, but most often the Kramer version is used. Kramer derived two models for $f(h)$, using two regimes. For the regime below the peak in $f(h)$ he assumed that flux motion primarily occurs by de-pinning of individual flux-lines, whereas in the regime above the peak in $f(h)$, de-pinning occurs by synchronous shear of the flux-line lattice around line pins which are too strong to be broken. The latter is referred to as the flux shear model and for this regime Kramer proposed~\cite{Kramer1973}:
\begin{equation}
F_{\rm p}  = \frac{{C_{66} }}{{12\pi ^2 \left( {1 - a_0 \sqrt \rho  } \right)^2 a_0 }}, \label{Kramerfp}
\end{equation}
where $C_{66}$ represents the elastic shear stiffness of the flux-line lattice, $a_0$ is the flux-line spacing and $\sqrt \rho$ represents the density of the pinning planes. In bulk Nb$_3$Sn, the grain boundaries are the main pinning centers~\cite{Scanlan1975,Shaw1976,Marken1986b,Schauer1981,West1977} and can thus be interpreted as the pinning planes yielding $1/\sqrt \rho=d_{\rm{av}}$, the average grain size.

One relation for the shear modulus of the flux-line lattice of a high field superconductor without paramagnetic limiting, was derived by Labusch~\cite{Labusch1969b}:
\begin{equation}
C_{66} \left( H \right) = 7.4 \times 10^4 \frac{{\left( {\mu _0 H_{\rm c2} } \right)^2 }}{{\kappa _1^2 }}\left( {1 - h} \right)^2 \ \left[ {{\rm{N/m}}^{\rm{2}} } \right]. \label{Labuschc66}
\end{equation}
The flux-line spacing can be estimated by assuming a triangular flux-line lattice~\cite{Tinkham1996}:
\begin{equation}
a_\vartriangle  \left( H \right) = \left( {\frac{4}{3}} \right)^{0.25} \left( {\frac{{\phi _0 }}{{\mu _0 H}}} \right)^{0.5}. \label{fluxlinespacing}
\end{equation}
Combining \eref{Kramerfp} with \eref{Labuschc66}, assuming $1/\sqrt \rho=d_{\rm{av}}$ and a triangular flux-line lattice with a spacing as given by \eref{fluxlinespacing} results in:
\begin{equation}
F_{\rm p} \left( H \right) = 12.8\frac{{\left( {\mu _0 H_{\rm c2} } \right)^{2.5} }}{{\kappa _1^2 }}\frac{{h^{0.5} \left( {1 - h} \right)^2 }}{{\left( {1 - {{a_\vartriangle  \left( H \right)} \mathord{\left/ {\vphantom {{a_\vartriangle  \left( H \right)} {d_{\rm{av}} }}} \right. \kern-\nulldelimiterspace} {d_{\rm{av}} }}} \right)^2 }}\ \left[ {{\rm{GN/m}}^{\rm{3}} } \right]. \label{kramerpinning}
\end{equation}
Using the Kramer form therefore results in $\nu=2.5$ and $\gamma=2$ in \eref{fietzwebb}, if $f(h)$ is defined as $f(h)=h^{0.5}(1-h)^2$ and $(1-a_\vartriangle(H)/d_{\rm{av}}) \cong 1$, which will be validated below.  Many authors indeed state $\nu=2.5$~\cite{Summers1991,Kramer1973,Dew-Hughes1974,Cooley2004,Ekin1980,Hampshire1985}.

Higher values for $\nu$ can be found if experimental results are described with non-Kramer-like pinning. Deviating values for $\nu$ in \eref{fietzwebb} are reported by Kroeger \textsl{et al.}~\cite{Kroeger1980} who claim $\nu=3.0$ and by Cheggour and Hampshire~\cite{Cheggour2002} who state $\nu=3.2$. It should be noted that both studies neglect $\kappa_1$ and fit the experimental results to $F_{\rm p}\propto(1-h)^3$ and $F_{\rm p}\propto(1-h)^{3.5}$ respectively. This choice suggests different pinning behavior. It will be explained below that such high values for the power above $(1-h)$ can arise when sample inhomogeneities are ignored. Here, they will be accounted for, which results in the experimentally supported assumption that $F_{\rm p}\propto(1-h)^2$, as will be discussed in more detail in Section \ref{Fp/Fpmax}.

Combining \eref{kramerpinning} with the reverse definition of the Lorentz force gives (Suenaga and Welch in~\cite{Suenaga1980}):
\begin{equation}
J_{\rm c}^{0.5} \left( {\mu _0 H} \right)^{0.25}  = \frac{{1.1 \times 10^5 }} {{\kappa _1 }}\frac{{\mu _0 \left( {H_{\rm c2}  - H} \right)}} {{\left( {1 - {{a_\vartriangle  \left( H \right)} \mathord{\left/  {\vphantom {{a_\vartriangle  \left( H \right)} {d_{\rm{av}} }}} \right.  \kern-\nulldelimiterspace} {d_{\rm{av}} }}} \right)}}.
\end{equation}

In most technical wires $d_{\rm{av}}$ is 100 to 200 nm, whereas the flux-line spacing decreases rapidly below 50 nm for magnetic fields above 1 T. Therefore $a_\vartriangle(H)/d_{\rm{av}} \ll 1$ and $J_{\rm c}^{0.5}(\mu_0H)^{0.25}$ will approximately be linear in $H$ and can be referred to as a Kramer function $f_{\rm K}$:
\begin{equation}
 \label{kramerplot}
 f_{\rm K} \left( H \right) \equiv J_{\rm c}^{0.5} \left( {\mu _0 H} \right)^{0.25}  \cong \frac{{10^5 }}{{\kappa _1 }}\mu _0 \left( {H_{\rm c2}  - H} \right).
\end{equation}
A linear extrapolation of \eref{kramerplot} to $J_{\rm c}=0$ then yields $H_{\rm c2}$ and the slope can be used to calculate $\kappa_1$. The Kramer extrapolated critical field will be referred to as $H_{\rm c2}^{\rm K}$ to distinguish it from a measured $H_{\rm c2}$.

The pinning force, as given by \eref{kramerpinning}, peaks at $h=0.2$ if $a_\vartriangle(H) \ll d_{\rm{av}}$ to yield the maximum pinning force:
\begin{equation}
F_{\rm p \max }  = \frac{{3.7\left( {\mu _0 H_{\rm c2} } \right)^{2.5} }}{{\kappa _1^2 }}\ \left[ {{\rm{GN/m}}^{\rm{3}} } \right].
\end{equation}
\subsubsection{Low field linearity of Kramer plots}
The dependence $F_{\rm p}\propto(1-h)^2$ occurs when shear deformation of the flux-line lattice is the primary de-pinning mechanism as was mentioned above. This was shown experimentally by Cooley \textsl{et al.}~\cite{Cooley2002} to be the case when the grain size is substantially larger than the flux-line spacing $ \left( d_{\rm{av}}>2a_\vartriangle(H)\right)$. If the grain size is comparable to the flux-line spacing, a direct summation of individual pinning interactions between grain boundaries and flux-lines occurs, which corresponds to $F_{\rm p}\propto(1-h)$~\cite{Labusch1969b,Campbell1972,Cooley2002}. Plots of the Kramer function $f_{\rm K}$ from \eref{kramerplot} versus magnetic field (a `Kramer plot'), calculated from $J_{\rm c}$ data measured on Nb$_3$Sn conductors, are generally linear in the `standard' $J_{\rm c}$ measurement regime of about $0.2<h<0.6$. This supports shearing of the flux-line lattice as the primary de-pinning mechanism, provided that the grain size is larger than about 100 nm~\cite{Cooley2002}. Moreover, Kramer plots resulting from magnetization data on Powder-in-Tube processed wires, reacted at various temperature and time combinations, were shown to be perfectly straight down to $h=0.07$ at various temperatures by Fischer~\cite{Fischer2002a}. Also transport $J_{\rm c}$ measurements on different bronze processed wires indicated straight Kramer plots at various temperatures down to $h=0.04$~\cite{Godeke2001,Hampshire1985}. Observed downward curvature when approaching lower magnetic fields can, apart from a different pinning behavior, additionally arise through self-fields generated by the large transport current densities. Accurate correction for these self-fields is non-trivial.

These observations validate the application of \eref{kramerpinning} to very low fields and support the assumption that the $(1-a_\vartriangle/d_{\rm{av}})$ term has negligible influence and can be approximated by $(1-a_\vartriangle/d_{\rm{av}})\cong1$. Moreover, they indicate that for technical wires, which have an average grain size above 100 nm, shear deformation of the flux-line lattice apparently remains the primary de-pinning mechanism also below the maximum in $F_{\rm p}$. The assumptions used to arrive at the Kramer model have received some criticism~\cite{Brandt1977}. In practice, however, using the high magnetic field description of Kramer appears to hold for Nb$_3$Sn wires, at least down to 1 T.
\subsubsection{Linearity of Kramer plots approaching $H_{\rm{c2}}$}
Above $h\cong0.8$, non-linearities are mostly observed in Kramer plots of wire results when approaching the maximum $H_{\rm c2}$ that is present in the wire. The pinning force function \eref{kramerpinning}, combined with the approximation $(1-a_\vartriangle(H)/d_{\rm{av}})\cong 1$ is therefore often written in a more general form:
\begin{equation}
\label{pinningfunction}
F_{\rm p} \left( H \right) = F_{\rm p \max }f(h)\cong C\frac{{\left( {\mu _0 H_{\rm c2} } \right)^\nu  }}{{\kappa _1^\gamma  }}h^p \left( {1 - h} \right)^q,
\end{equation}
where $C$ is a constant and $p$ and $q$ determine $f(h)$. This allows for insertion of different values of $p$ and $q$ that implicate a different pinning behavior, but most probably originate from inhomogeneity averaging. Such different values for $p$ and $q$ lead to still ongoing discussion in the literature whether or not a general pinning mechanism, or $f(h)$, can be assumed for Nb$_3$Sn wires.

\begin{figure}
\includegraphics [scale=1]{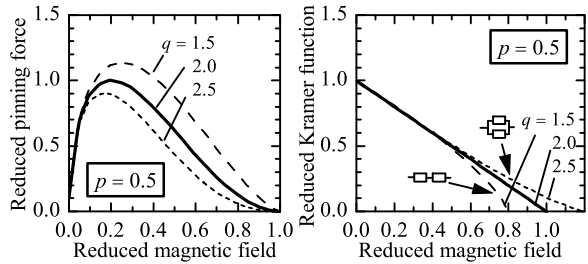}
\caption{\label{kramersim} Calculated changes with $q$ in the field dependence of the bulk pinning force (left plot) and Kramer function (right plot).}
\end{figure}

The field dependence of the bulk pinning force as given by \eref{pinningfunction} is shown in figure \ref{kramersim} and combined with Kramer plots for various values of $q$ around 2. A variation of $p$ only influences the low field region. The aforementioned observations of perfectly straight Kramer plots at lower magnetic fields for bronze and Powder-in-Tube processed wires, however, render strong deviations of $p$ from 0.5 unlikely.

A $\pm~25\%$ variation in $q$ results in a change in the maximum pinning force and its position and an up- or downward curvature in the Kramer plot for $h\rightarrow 1$. The effect of a lowering of $q$, i.e. an increase of the maximum pinning force combined with a shift of its position towards higher magnetic field is identical to what has been observed experimentally through grain refinement in Powder-in-Tube wires by Cooley \textsl{et al.}~\cite{Cooley2002}. Upward and downward tails in Kramer plots have both been observed experimentally (e.g. Ekin in~\cite{Matsushita1985},~\cite{Godeke2005b}, Suenaga in~\cite{Foner1981b},~\cite{Drost1984}) and have been attributed to microstructural~\cite{Flukiger1987} and compositional (Ekin in~\cite{Matsushita1985},~\cite{Godeke2005b}, Cooley \textsl{et al.}~\cite{Cooley2004}, Suenaga and Welch in~\cite{Suenaga1980}) origins. A range of experimentally detectable $H_{\rm{c}2}$ values could also be attributed to flux-creep and thermal activated flux-flow (e.g. the presence of an `irreversibility field' in high temperature superconductors), but we find that these effects are negligibly small compared to the severe influence of compositional inhomogeneities on $H_{\rm{c}2}$~\cite{Godeke2006}, especially at temperatures below 20~K.

The most convincing support for attributing curvature in Kramer plots when $h\rightarrow 1$ to inhomogeneities is given by Cooley \textsl{et al.}~\cite{Cooley2004}, who numerically modeled the filaments in a Powder-in-Tube processed wire as concentric shells with different Sn content and thus different $H_{\rm c2}$ and $T_{\rm c}$. It was assumed that each composition behaves according to $q=2$. An area weighted summation of the transport current density of such a system of parallel paths of variable property showed that a concave tail appears in the overall Kramer plots when the maximum available $H_{\rm c2}$ is approached. Increasing the severity of the Sn gradient in the A15 sections increases the occurrence of such tails in otherwise linear overall Kramer plots. This parallel path model is supported by measurements on similar PIT processed wires~\cite{Godeke2005a,Godeke2005b}. Such an area weighted summation over parallel paths of lesser and better A15 quality results in a similar overall Kramer plot as for $p=0.5$ and $q=2.5$ in figure \ref{kramersim}. A downward tail in Kramer plots, as is often observed in bronze processed wires, can be described with $q<2$ as is shown in figure \ref{kramersim}, thereby again implicitly assuming a different de-pinning mechanism. However, assuming a serial connection of lesser and better A15 quality regions for the more irregular filaments in bronze processed wires, also results in a downward tail in Kramer plots when $h\rightarrow 1$, since the lower quality A15 sections will determine the overall transport properties. This argumentation is a strong indication that the curvature is not due to different de-pinning mechanisms but to inhomogeneity effects and thus that $q$ should be 2.

It can thus be stated that the Kramer flux-line shear model, at constant composition, is valid over nearly the full magnetic field range, i.e. up to $H_{\rm c2}$ and down to a magnetic field where the flux-line lattice spacing becomes approximately more than half of the A15 grain size~\cite{Cooley2002}, which is below 1 T from \eref{fluxlinespacing}. Non-linearities in Kramer plots when approaching the maximum present $H_{\rm c2}$ are thus attributed to inhomogeneity averaging. The effective bulk values $H_{\rm c2}^*$ and $T_{\rm c}^*$, as detected by critical current measurements, are weighted averages over the A15 volume, but it is not always clear how the averaging occurs. The discussion above leads to the conclusion that there is no \textsl{a priori} reason to deviate from $F_{\rm p} \propto h^{0.5}(1-h)^2$, provided that the inevitable presence of A15 inhomogeneities in wires is recognized.
\subsection{\label{Hc2-T} Temperature dependence}
To come to an overall description for $J_{\rm c}(H,T,\epsilon)$, the temperature dependence of the bulk pinning force is considered. In observing \eref{pinningfunction} it is clear that temperature dependence occurs through temperature dependence of $H_{\rm c2}$ and $\kappa_1$:
\begin{equation} \label{fpht}
F_{\rm p} \left( {H,T} \right) \cong C\frac{{\left[ {\mu _0 H_{\rm c2} \left( T \right)} \right]^\nu  }}{{\kappa _1 \left( T \right)^\gamma  }}f\left( h \right),
\end{equation}
with $h=H/H_{\rm c2}(T)$. It is thus required to discuss the available descriptions for $H_{\rm c2}(T)$ and $\kappa_1(T)$. For both parameters, empirical as well as microscopic alternatives are available. In the Summers relation~\cite{Summers1991} empirical forms are used. It will be shown here, how these can be replaced by alternatives with a better connection to the microscopic theory.
\subsubsection{\label{Summershc2} Empirical temperature dependence}
First, the empirical temperature dependence of the upper critical field, as used in the Summers relation, is introduced.  Based on earlier work by Suenaga in~\cite{Foner1981b} and Hampshire \textsl{et al.}~\cite{Hampshire1985}, Summers \textsl{et al.}~\cite{Summers1991} proposed an empirical relation for the temperature dependence of the upper critical field, thereby improving the correspondence between the available scaling relations and measured results:
{\setlength\arraycolsep{1pt}
\begin{eqnarray}
\label{summershc2t}
\frac{{H_{\rm c2} \left( T \right)}}{{H_{\rm c2} \left( 0 \right)}} & = & \left( {1 - t^2 } \right)\frac{{\kappa _1 \left( T \right)}}{{\kappa _1 \left( 0 \right)}} \nonumber \\
& = & \left( {1 - t^2 } \right)\left[ {1 - 0.31t^2 \left( {1 - 1.77\ln t} \right)} \right],
\end{eqnarray}}
where $t=T/T_{\rm c}$. In this relation an empirical fit for $\kappa_1(T)/\kappa_1(0)$ is inserted in $\kappa_1=H_{\rm c2}/(\surd2H_{\rm c})$ and $H_{\rm c}(t)/H_{\rm c}(0)=(1-t^2)$ is used for the temperature dependence of the thermodynamic critical field. It has to be emphasized that Hampshire \textsl{et al.}~\cite{Hampshire1985} claim a validity range only below 13.5 K for their calculated $\kappa_1(T)$ values.
\subsubsection{\label{mdgt} Microscopic based $H_{\rm c2}(T)$}
An early microscopic description for the temperature dependence of the upper critical field was simultaneously derived in 1964 by Maki~\cite{Parks1969,Maki1964} and De Gennes~\cite{deGennes1966,deGennes1964} which, written in the De Gennes form yields the implicit relation (MDG relation):
\begin{equation}
\label{MDG}
\ln \left( {\frac{T}{{T_{\rm c} \left( 0 \right)}}} \right) = \psi \left( {\frac{1}{2}} \right) - \psi \left( {\frac{1}{2} + \frac{{\hbar D\mu _0 H_{\rm c2} \left( T \right)}}{{2\phi _0 k_{\rm B} T}}} \right).
\end{equation}
The function uses only two parameters, namely $T_{\rm c}(0)$ (which can be measured) and the diffusion constant of the normal conducting electrons $D$. The diffusion constant is inversely proportional to the slope of the $H_{\rm c2}(T)$ dependence at $T_{\rm c}(0)$~\cite{Godeke2005a,Godeke2005b}. The other parameters are the reduced Planck constant ($\hbar$), the magnetic flux quantum ($\phi _0$) and the Boltzmann constant ($k_{\rm B}$). The terms  $\psi(x)$ represent the digamma function~\cite{Abramowitz1965}. The description was derived assuming a dirty superconductor (i.e. $\ell \ll \xi$), it uses a spherical Fermi surface approximation and assumes a constant density of states at the Fermi level $N(E_{\rm F})$ and a weak electron-phonon interaction. Furthermore it assumes no paramagnetic limitation of $H_{\rm c2}(T)$ and an absence of spin-orbit scattering. A normalized form of \eref{MDG} with respect to temperature can be defined:
\begin{equation}\label{MDGshort}
MDG\left( t \right) \equiv \frac{{H_{\rm c2} \left( t \right)_{\rm {MDG}} }}{{H_{\rm c2} \left( 0 \right)_{\rm{MDG}} }},
\end{equation}
in which $H_{\rm c2}(t)_{\rm{MDG}}$ represents $H_{\rm c2}(T)$ calculated using \eref{MDG}.

It can be convenient for practical applications to avoid the need to solve the implicit MDG relation by replacing it with an explicit expression. An approximation of \eref{MDG} for the entire temperature range yields:
\begin{equation}\label{approxmdg}
    MDG\left( t \right) \cong \left( {1 - t^{1.52} } \right).
\end{equation}
This power of 1.52 is practically identical to the value of 1.5 which was found from a similar fit as \eref{approxmdg} to (partly) extrapolated $H_{\rm c2}^*(T)$ results by Cheggour and Hampshire~\cite{Cheggour2002}. A power of 1.5 was also found recently from analysis using Eliashberg theory by Oh and Kim~\cite{Oh2005a}.

Application of the full Eliashberg based formalisms is the most correct approach, since it accounts for all the electron-phonon interaction strengths. However, there are two reasons for applying only the simplest form of the microscopic descriptions [i.e. \eref{MDG}] to wire results. First, since wires are inherently inhomogeneous there is no single $H_{\rm c2}(T)$ but a distribution of properties~\cite{Godeke2005b}. Also, the inaccessibility of the normal state resistivity (which is a prime parameter in the microscopic theory) of the A15 in wires renders a proper connection to the theory practically impossible. Second, the use of the formal descriptions results in an increased number of parameters compared to \eref{MDG}. This can be avoided by implementing the additional parameters that are required for a proper connection to the microscopic theory in $D$ [and thus in the slope at $T_{\rm c}(0)$~\cite{Godeke2005b}] and $T_{\rm c}(0)$ in \eref{MDG}. To allow this it has to be assumed that the shape of $H_{\rm c2}(T)$, at least to first order, remains unchanged.

The simplest form of the microscopic descriptions, the MDG relation \eref{MDG}, was tested previously and found to fit $H_{\rm c2}(T)$ for all available literature Nb$_3$Sn results and measurements on a multitude of wires with very high accuracy~\cite{Godeke2005b}. This accuracy holds for the Orlando \textsl{et al.} thin film results~\cite{Orlando1979} at all Sn concentrations, the Foner and McNiff single- and polycrystalline results~\cite{Foner1976,Foner1981a}, and the Jewell \textsl{et al.} bulk results~\cite{Jewell2004}. The single crystals are close to stoichiometric and thus strong coupling~\cite{Moore1979} and most probably clean. The thin films and bulk samples are of varying degree of resistivity and Sn content and cover therefore a large range from approximately clean to dirty and weak to strong coupling. This leads to the conclusion that the shape of the field-temperature phase boundary is, within the experimental error bars, independent of the electron-phonon coupling strength (which can also be confirmed from the theory~\cite{Rainer1973}), independent on whether the material is clean or dirty, and not influenced by the (unlikely~\cite{Schossmann1986,Orlando1979,Godeke2005a,Beasley1982}) presence of PPL and spin-orbit scattering. It can thus be expected that it can be validated for the entire range of compositions that are present in wires. This is confirmed by the fact that it also holds for Kramer extrapolated, inhomogeneity averaged $H_{\rm c2}^{\rm K}(T)$ results~\cite{Godeke2005b}.
\subsubsection{\label{microkappa} Microscopic based $\kappa_1(T)$}
From the microscopic theory~\cite{Rainer1974} it follows that for weak coupling superconductors in the dirty limit $\kappa_1(0)/\kappa_1(T_{\rm c})=1.2$. In the strong coupling limit this ratio saturates at about 1.5~\cite{Rainer1974}, which means that the maximum expectable change in $\kappa_1(T)$ is 50\% from 0 K to $T_{\rm c}$.

For calculation of $\kappa_1(T)/\kappa_1(0)$ for a given interaction strength the nonlinear Eliashberg equations~\cite{Eliashberg1960} will have to be solved as was done numerically by Rainer and Bergmann~\cite{Rainer1974}. No general simple function for $\kappa_1(T)/\kappa_1(0)$, derived directly from the electron-phonon spectrum and thus valid for all $\lambda_{\rm{ep}}$, can therefore be given for wires. Fortunately, a generalized function can be derived for $\kappa_1(T)/\kappa_1(0)$ through $\kappa_1=H_{\rm c2}/(\surd2H_{\rm c})$, using the microscopic form for $H_{\rm c2}(T)/H_{\rm c2}(0)$ \eref{MDG} combined with the temperature dependence of $H_{\rm c}$.

It is well accepted that $H_{\rm c}(t)/H_{\rm c}(0)$ is for all superconductors very close to $(1-t^2)$~\cite{Tinkham1996}. Recent calculations based on Eliashberg theory yield $H_{\rm c}(t)/H_{\rm c}(0)=(1-t^{2.17})$~\cite{Oh2005a}. The deviation from $(1-t^2)$ is measured for Nb$_3$Sn~\cite{Guritanu2004} and is within about 2\%. A fit to the measured $H_{\rm c}(t)/H_{\rm c}(0)$ yields $(1-t^{2.07})$. For practical applications a power of 2 will thus be sufficiently accurate. Combining \eref{MDG}/\eref{MDGshort} or \eref{approxmdg} with $\kappa_1=H_{\rm c2}/(\surd2H_{\rm c})$ and $H_{\rm c}(t)/H_{\rm c}(0)=(1-t^2)$ thus yields:
\begin{equation}\label{kappat}
    k\left( t \right) = \frac{{\kappa _1 \left( t \right)}}{{\kappa _1 \left( 0 \right)}} = \frac{{MDG\left( t \right)}}{{1 - t^2 }} \cong \frac{{1 - t^{1.52} }}{{1 - t^2 }}.
\end{equation}
By implementing an accurate microscopic form for $H_{\rm c2}(T)$ in the scaling relations also $\kappa_1(T)$ is accurately known, rendering an empirical version redundant.
\subsection{\label{Strain} Strain dependence}
\subsubsection{Strain dependence in Nb$_3$Sn wires}
When a Nb$_3$Sn composite wire is subjected to longitudinal compressive strain, the critical current density, upper critical field and critical temperature reduce approximately linearly and reversibly with strain. When a wire is subjected to a longitudinal tensile strain, the critical parameters increase approximately proportionally and reversibly with strain until a parabolic-like peak is reached, after which the properties reduce approximately linearly with strain. The larger thermal contraction of the matrix materials compared to the thermal contraction of the Nb$_3$Sn results in an axial pre-compression of the A15 when cooled from reaction temperature to the test temperature below 20~K. When the wire is axially loaded after cool-down, this pre-compression is minimized and $J_{\rm c}$, $H_{\rm c2}$ and $T_{\rm c}$ increase. The position of the maximum in the strain dependency curve appears at the point where the three-dimensional deviatoric strain components in the A15 are minimal. In practice this correlates closely to the point where the axial pre-compression of the A15 is minimal. The axial pre-compression is mostly identified by $\epsilon_{\rm m}$, whereas the position of the minimum in the deviatoric strain components is identified as $\delta$. Several models have been introduced in the literature to describe this behavior, ranging from polynomial fits to relations based on the Eliashberg theory. These will be discussed below.

When an A15 lattice is deformed, its vibration modes will change. In addition, the electronic structure will be modified and therefore $N(E_{\rm F})$. It is thus reasonable to expect a strain induced variation of the electron-phonon interaction spectrum and the phonon density of states and hence, a change in the interaction constant $\lambda_{\rm{ep}}$. The difference between the strain sensitivity of $H_{\rm c2}$ and $T_{\rm c}$ represents an important link to microscopic understanding since both depend differently on $N(E_{\rm F})$ and $\lambda_{\rm{ep}}$. It is found experimentally that~\cite{Welch1980,Ekin1980}:
\begin{equation}\label{straindep4K}
    \frac{{H_{\rm c2} \left( {4.2K,\epsilon } \right)}}{{H_{\rm{c2m}} \left( {4.2K} \right)}} \cong \left( {\frac{{T_{\rm c} \left( \epsilon  \right)}}{{T_{\rm{cm}} }}} \right)^\varpi,
\end{equation}
where the index `${\rm m}$' indicates the values of $H_{\rm c2}$ and $T_{\rm c}$ at the maximum of the axial strain dependency curve. The power $\varpi\cong3$, indicates that $H_{\rm c2}$ is roughly 3 times more sensitive to strain than $T_{\rm c}$. This was found to be valid at all temperatures (e.g.~\cite{Godeke2003,tenhaken1999}). An attempt was made by Welch~\cite{Welch1980}, following earlier work by Testardi~\cite{Testardi1971,Testardi1973,Testardi1972}, to explain the observed difference in strain sensitivity on the basis of the strong coupling renormalized BCS theory, using a McMillan~\cite{McMillan1968} based, Allan and Dynes~\cite{Allen1975} strong coupling formulation for $T_{\rm c}$. It was concluded that $\varpi\cong3$ cannot be explained by either a change in $N(E_{\rm F})$ or $\lambda_{\rm{ep}}$ alone and the strain dependence should therefore be described by a combination of both. It should be pointed out, however, that McMillan based descriptions for $T_{\rm c}$ are valid for $\lambda_{\rm{ep}}<1.5$~\cite{Kresin1993}, whereas reported values for Nb$_3$Sn are mostly higher~\cite{Orlando1979}. It is clear, however, that improved understanding of the value for $\varpi$ should result from exact calculations of strain induced modifications on the full electron-phonon interaction spectrum and the density of states. This might be possible through a combination of the recent efforts of Markiewicz~\cite{Markiewicz2004a,Markiewicz2004b,Markiewicz2004c} and Oh and Kim~\cite{Oh2005a,Oh2005b}. An important conclusion drawn by Welch is that in a three-dimensional strain description the deviatoric strain components dominate the strain sensitivity of Nb$_3$Sn and that hydrostatic components have, in comparison, a negligible effect.

Most strain dependencies are defined through the strain dependence of the upper critical field or the critical temperature. Two general strain dependent terms will therefore be defined:
\begin{equation}\label{seps}
    s\left( \epsilon  \right) \equiv \frac{{H_{\rm c2} \left( \epsilon  \right)}}{{H_{\rm{c2m}} }}\equiv \left( \frac{{T_{\rm c} \left( \epsilon  \right)}}{{T_{\rm{cm}} }} \right)^\varpi,
\end{equation}
under the assumption that $s(\epsilon)$ is independent of temperature and $\varpi\cong3$.
\subsubsection{Available models}
Five models describing the strain sensitivity of Nb$_3$Sn can be distinguished throughout the literature. The simplest form is an introduction of hydrostatic strain though a pressure term. Although this yields a relatively simple connection to thermodynamic calculations to describe single crystal hydrostatic experiments, it is not suited for more practical systems such as wires due to the dominating deviatoric components. An empirical fit for axial deformations in wires was recently adapted by Hampshire and co-authors~\cite{Keys2003} in the form of a fourth-order polynomial. Although this will obviously result in accurate fits to experimental data, it lacks any connection to the underlying physics, and is thus not suited for the goals set out for this article. The three remaining models that have been proposed to describe experiments on wires will be discussed next in chronological order.

The first model that was proposed to describe the strain dependence of the critical properties of Nb$_3$Sn composite wires was the so-called power-law model, introduced by Ekin in 1980~\cite{Ekin1980}. The model was developed  based on measurements using an axial pull strain device. The investigated compressive strain range was therefore limited to the thermal pre-compression of the A15, introduced by the matrix components of the wires. A power-law dependence was proposed for $s(\epsilon)$ using the extrapolated upper critical field~\cite{Ekin1980}. In this power-law dependence, a different axial strain ($\epsilon_{\rm a})$ sensitivity is used for  $\epsilon _{\rm a}<0$ and for $\epsilon _{\rm a}>0$. The power-law model is an empirical description that does not account for the three-dimensional nature of strain, but includes the generally observed asymmetry in axial strain experiments. Observable linearities at large compressive longitudinal strains are not accounted for and it is therefore valid over a limited strain regime.

The second model that was proposed was the so-called deviatoric strain model. For high compressive strains Ten Haken found an approximate linear dependence of $H_{\rm c2}^{\rm K}$ on axial strain, deviating from the power-law behavior~\cite{tenhaken1995a,tenhaken1996}. These deviations from the power-law model at large compressive axial strain values, combined with results of experiments on quasi two-dimensional Nb$_3$Sn tape conductors~\cite{tenhaken1994,tenhaken1995b,tenhaken1997} (which allow an analytical calculation of the three-dimensional strain components) led to an alternative description based on the non-hydrostatic, i.e. distortional or deviatoric strain~\cite{tenhaken1994,Godeke2002b}. In this deviatoric strain model only the second strain invariant is considered. The model is empirical, does account for the three-dimensional nature of strain and linearity at large compressive axial strains, but is symmetrical around the maximum in the strain dependency curve, and is claimed to be valid only for the compressive strain regime.

The third model available in the literature is the so-called full invariant strain analysis~\cite{Markiewicz2004a,Markiewicz2004b,Markiewicz2004c}. This model is based on the assumption that the strain dependence of $T_{\rm c}$ results from the strain dependence of the phonon modes in the Nb$_3$Sn crystal lattice. The model uses a full invariant strain analysis which correlates the strain dependence of $T_{\rm c}$ to strain induced changes in the phonon frequency spectrum using relations that stem from Eliashberg theory and a strain energy potential function. The main result of the full invariant model is that it is able to demonstrate the effect of the harmonic and anharmonic terms in the strain energy potential function independently. It explains how the $T_{\rm c}$ reduction due to hydrostatic strain decreases linearly with axial strain. It further explains the experimentally observed parabolic-like behavior around the maximum, the linearity for large compressive axial stain values and the asymmetry, in terms of the second and third strain invariants. For now, unfortunately, it does not include a description for $H_{\rm c2}^*(\epsilon)$, which is required for $J_{\rm{c}}$ scaling. The latter might be provided by a combination of recent promising work by Oh and Kim~\cite{Oh2005a,Oh2005b} and a full invariant analysis.
\subsubsection{Empirical correction to the deviatoric strain model}
The full invariant analysis is the most proper method available for the introduction of strain dependency of the critical properties. The resulting relations, however, remain somewhat complex compared to the empirical fits, and a complete description for $T_{\rm c}(\epsilon)$ \emph{and} $H_{\rm c2}^*(\epsilon)$ is not fully developed yet. It is clear that the empirical models all lack specific details in comparison to the full invariant analysis. An empirical model that accounts for three-dimensionality, linearity at large axial strain, and asymmetry, in agreement with the invariant analysis, can be obtained by assuming a simple mathematical correction to the Ten Haken deviatoric strain model~\cite{Godeke2005a}. Our starting point is relation (3) in~\cite{Godeke2002b}:
\begin{equation}\label{benniedev}
    \mu _0 H_{\rm c2}^{\rm K} \left( {\epsilon_{\rm{dev}} } \right) \cong \mu _0 H_{\rm c2}^{\rm K} \left( 0 \right) - C_{\rm{dev}} \sqrt {\left( {\epsilon_{\rm{dev}} } \right)^2  + \left( {\epsilon _{0,\rm d} } \right)^2 } ,
\end{equation}
in which $\epsilon_{\rm{dev}}$ represents the second deviatoric strain invariant, $C_{\rm{dev}}$ represents the slope in the linear regime of $\mu_0H_{\rm c2}^{\rm K}(\epsilon_{\rm{dev}})$, and $\epsilon _{0,\rm d}$ represents the non-axial, remaining strain components when the axial strain is zero. In axial form \eref{benniedev} can be written as~\cite{Godeke2002b,Godeke2005a}:
\begin{equation}\label{bennieaxial}
    \mu _0 H_{\rm c2}^{\rm K} \left( {\epsilon_{\rm a} } \right) \cong \mu _0 H_{\rm c2}^{\rm K} \left( 0 \right) - C_{\rm a}^{\rm '} \sqrt {\left( {\epsilon_{\rm a} } \right)^2  + \left( {\epsilon _{0,\rm a} } \right)^2 },
\end{equation}
where $C_{\rm a}^{\rm '}$ is a constant and $\epsilon_{0,\rm a}$ represents the remaining strain components, as with the factor $\epsilon _{0,\rm d}$ in \eref{benniedev}. To account, in the axial form of the Ten Haken model, for a stronger reduction at tensile strains, a linear term causing a reduction of $H_{\rm c2}^{\rm K}$ with axial strain is required. This can be achieved by including an additional linear term in \eref{benniedev}:
{\setlength\arraycolsep{1pt}
\begin{eqnarray}\label{arnodev}
    \mu_0 H_{\rm c2}^{\rm K} \left( {\epsilon_{\rm{dev}} } \right) \cong \mu _0 H_{\rm c2}^{\rm K} \left( 0 \right) &-& C_{\rm{inv}2} \sqrt {\left( {\epsilon_{\rm{dev}} } \right)^2  + \left( {\epsilon _{0,\rm d} } \right)^2 }   \nonumber \\
    &-& C_{\rm{inv}3} \epsilon_{\rm{dev}},
\end{eqnarray}}
where $C_{\rm{inv}2}$ and $C_{\rm{inv}3}$ are constants representing the second and third strain invariants. The term $\epsilon _{0,\rm d}$ now represents the initial suppression of $H_{\rm c2}^{\rm K}$ due to hydrostatic strain. A similar route can be followed for the axial version \eref{bennieaxial}, resulting in:
{\setlength\arraycolsep{1pt}
\begin{equation}\label{arno1}
    \mu_0 H_{\rm c2}^{\rm K} \left( {\epsilon_{\rm a} } \right) = \mu_0 H_{\rm c2}^{\rm K} \left( 0 \right) - C_{\rm a1}^{\rm '} \sqrt {\left( {\epsilon _{\rm a} } \right)^2  + \left( {\epsilon _{0,\rm a} } \right)^2 }- C_{\rm a2}^{\rm '} \epsilon_{\rm a},
\end{equation}}
in which $\epsilon_{\rm a}=\epsilon_{\rm{applied}}+\delta$. The correction $C_{\rm a2}^{\rm '}\epsilon_{\rm a}$ effectively rotates the strain dependency curve around its maximum. This causes the axial position of the maximum to shift by a factor $\epsilon_{\rm{sh}}= C_{\rm a2}^{\rm '}\epsilon_{0,\rm a}/[(C_{\rm a1}^{\rm '})^2-(C_{\rm a2}^{\rm '})^2]^{0.5}$. For large asymmetry, $\delta$ (the axial strain at which the minimum in the deviatoric strain occurs) will then deviate significantly from the observable axial position of the maximum $\epsilon_{\rm m}$. In axial strain experiments it is more convenient to normalize the strain dependence function to the observable maximum, also to avoid the introduction of an additional parameter through the use of $\delta$ in combination with $\epsilon_{\rm m}$. The rotation around the maximum also causes the normalized form \eref{seps} to increase above 1 at the position of the maximum. This is undesirable since in scaling relations the reduction in e.g. $H_{\rm c2}^{\rm K}$ is usually normalized to the maximum observable value at the peak of the measured strain dependency curve, and not to the `strain free' value since this is not accessible in an axial strain experiment. This can be prevented by re-normalizing $s(\epsilon)$ to the maximum observable critical field value. A re-normalization of $s(\epsilon)$ to the observable peak position and peak value in an axial strain experiment leads to a new axial form:
{\setlength\arraycolsep{1pt}
\begin{eqnarray}\label{sarno}
    s\left( {\epsilon_{\rm a} } \right) &=& \frac{1}{{1 - C_{\rm a1} \epsilon _{0,\rm a} }} \biggl({C_{\rm a1} } \biggl[ \sqrt {\left( {\epsilon _{\rm{sh}} } \right)^2  +    \left( {\epsilon _{0,\rm a} } \right)^2 }     \nonumber \\
    &-& \sqrt {\left( {\epsilon_{\rm a}   -  \epsilon _{\rm{sh}} } \right)^2  + \left( {\epsilon _{0,\rm a} } \right)^2 }  \biggr] - C_{\rm a2} \epsilon_{\rm a}  \biggr) + 1, \nonumber \\
    \epsilon _{\rm{sh}}  &=& \frac{{C_{\rm a2} \epsilon _{0,\rm a} }}{{\sqrt {\left( {C_{\rm a1} } \right)^2  - \left( {C_{\rm a2} } \right)^2 } }}, \nonumber \\
    \epsilon_{\rm a}  &=& \epsilon_{\rm{applied}}  + \epsilon _{\rm m} .
\end{eqnarray}}
It has to be emphasized that this is an empirical correction to include the effects of the third strain invariant which, as suggested by the full invariant analysis, should be accounted for to describe asymmetry. It results in a simple description with three parameters (plus pre-strain) that accounts for the overall behavior as expected on the basis of a full invariant analysis and experimental observations. The hydrostatic strain components are included in the remaining strain term  $\epsilon_{0,\rm a}$ and the strain induced behavior resulting from the second and third invariants is accounted for through $C_{\rm a1}$ and $C_{\rm a2}$ respectively. Relation~(\ref{sarno}) will be used through the remainder of this article since it is the simplest form that still accounts for the three-dimensional nature of strain and is based on an underlying physics model. Possible future, more fundamental based, strain descriptions can be easily implemented in the general scaling relation, since all strain dependence is included explicitly in $s(\epsilon)$ in terms of a change in $H_{\rm{c}2}$ and/or $T_{\rm{c}}$ \eref{seps}.
\subsection{\label{F-HTeps} Field, temperature and strain dependence of the bulk pinning force}
The strain dependence described above has to be included in the relation for the magnetic field and temperature dependent bulk pinning force \eref{fpht} to arrive at a relation that can be used to inter- and extrapolate critical current measurements on wires. This requires the use of a strain dependent critical temperature as defined earlier by \eref{seps} and a temperature and strain dependent GL parameter and upper critical field:
\begin{equation}\label{hc2e}
    H_{\rm c2} \left( {T,\epsilon } \right) = H_{\rm{c2m}} \left( T \right)s\left( \epsilon  \right).
\end{equation}
Implementing this in the Kramer form of the field dependence of the bulk pinning force, i.e. \eref{kramerpinning}, and assuming $(1-a_\vartriangle/d_{\rm{av}})\cong1$ yields:
\begin{equation} \label{fphte}
    F_{\rm p} \left( {H,T,\epsilon } \right) \cong C\frac{{\left[ {\mu _0 H_{\rm c2} \left( {T,\epsilon } \right)} \right]^{2.5} }}{{\kappa _1 \left( {T,\epsilon } \right)^2 }}h^{0.5} \left( {1 - h} \right)^2 .
\end{equation}
The upper critical field thus delivers a term $s(\epsilon)^{2.5}$ acting directly on the bulk pinning force in addition to strain dependence arising through $f(h)$ and $T_{\rm c}(\epsilon)$. The strain dependence of $\kappa_1$ is, however, undefined since this requires knowledge of the strain dependence of $H_{\rm c}$ which can differ from $H_{\rm c2}(\epsilon)$. In addition, some strain dependence through strain induced changes of the flux-line to lattice interactions could be assumed. To arrive at an overall description, it is thus required to have knowledge of $\kappa_1(T,\epsilon)$, and/or to introduce a strain dependent constant $C(\epsilon)$.
\subsubsection{Ekin's unification of strain and temperature dependence}
In analyzing critical current versus strain data at 4.2 K Ekin~\cite{Ekin1980} found that the bulk pinning force scales, in analogy to the Fietz and Webb scaling law \eref{fietzwebb}, as:
\begin{equation}\label{ekin1}
    F_{\rm p} \left( {H,\epsilon } \right)_{4.2\;\rm K}  \propto \left[ {H_{\rm c2}^* \left( \epsilon  \right)} \right]^n f\left( h \right)
\end{equation}
where $n=1.0\pm0.3$ and the star again indicates a bulk average. The power of 1 differs from the power of 2.5 that is required for magnetic field and temperature scaling, as seen from \eref{fpht}. Ekin solved this inconsistency by postulating a combination in terms of explicit dependencies on temperature and strain. Neglecting the temperature and strain dependence of $\kappa_1$ he proposed:
{\setlength\arraycolsep{1pt}
\begin{eqnarray} \label{ekin2}
  F_{\rm p} \left( {H,T,\epsilon } \right) &=& Cs\left( \epsilon  \right)^n \left[ {\frac{{H_{\rm c2}^* \left( {T,\epsilon } \right)}}{{H_{\rm c2}^* \left( {0,\epsilon } \right)}}} \right]^\nu  f\left( h \right)  \nonumber \\
  &\cong& Cs\left( \epsilon  \right)^n \left( {1 - t^2 } \right)^\nu  f\left( h \right)
\end{eqnarray}}
where $h=H/H_{\rm c2}^*(T,\epsilon)$, $t=T/T_{\rm c}^*(\epsilon)$ (stars indicating bulk average values), $n\cong1$, $\nu\cong2.5$. Note that the third term only describes the normalized temperature dependence of the upper critical field approximated as $(1-t^2)$, in which strain sensitivity occurs only through $T_{\rm c}^*(\epsilon)$. The combination of strain and temperature sensitivity was introduced on a fully empirical basis, based on 4.2 K critical current data. This implicates, as was pointed out by Ekin, that the term $C/\kappa_1(T,\epsilon)^2$ in \eref{fphte} has to produce a strain dependency $s(\epsilon)^{1.5}$ to counteract the $s(\epsilon)^{2.5}$ term arising through the strain dependent upper critical field in \eref{fphte}. A similar magnetic field, temperature and strain dependence is used in various forms of generalized scaling relations which, by combining \eref{ekin2} with $F_{\rm p}= -J \times B$, result in an overall description for the critical current density in Nb$_3$Sn wires.
\subsubsection{Unification of the Kramer form}
The ratio $[\mu_0H_{\rm c2}(T,\epsilon)]^{2.5}/\kappa_1(T,\epsilon)^2$ in \eref{fphte} can be normalized to zero temperature to express the direct strain influence on $F_{\rm p}$ (i.e. not through $T_{\rm c}(\epsilon)$) separately, so that the result only contains temperature dependence and strain dependence through $T_{\rm c}(\epsilon)$. With the definitions:
{\setlength\arraycolsep{1pt}
\begin{eqnarray} \label{normalizedt}
    b\left( {T,T_{\rm c} \left( \epsilon  \right)} \right) &\equiv& \frac{{H_{\rm c2} \left( {T,\epsilon } \right)}}{{H_{\rm c2} \left( {0,\epsilon } \right)}}, \\
    k\left( {T,T_{\rm c} \left( \epsilon  \right)} \right) &\equiv& \frac{{\kappa _1 \left( {T,\epsilon } \right)}}{{\kappa _1 \left( {0,\epsilon } \right)}},
\end{eqnarray}}
this leads to:
{\setlength\arraycolsep{1pt}
\begin{eqnarray}
    F_{\rm p} \left( {H,T,\epsilon } \right) &=& \frac{{C\left[ {\mu _0 H_{\rm{c2m}} \left( 0 \right)} \right]^{2.5} }}{{\kappa _1 \left( {0,\epsilon } \right)^2 }}s\left( \epsilon  \right)^{2.5} \nonumber \\
     &\times& \frac{{b\left( {T,T_{\rm c} \left( \epsilon  \right)} \right)^{2.5} }}{{k\left( {T,T_{\rm c} \left( \epsilon  \right)} \right)^2 }}h^{0.5} \left( {1 - h} \right)^2.
\end{eqnarray}}
To retain consistency with Ekin's observation that $F_{\rm p}(\epsilon)_{4.2\ \rm{ K}}\propto s(\epsilon)$ (under the assumption that this is valid for all temperatures) it can for example be postulated that:
\begin{equation}
    \kappa _1 \left( {0,\epsilon } \right) = \kappa _{1\rm m} \left( 0 \right)s\left( \epsilon  \right)^{\sqrt {1.5} },
\end{equation}
leading to:
{\setlength\arraycolsep{1pt}
\begin{eqnarray}\label{kramerfphte}
F_{\rm p} \left( {H,T,\epsilon } \right) &=& \frac{{C\left[ {\mu _0 H_{\rm{c2m}} \left( 0 \right)} \right]^{2.5} }}{{\kappa _1 \left( 0 \right)^2 }}s\left( \epsilon  \right) \nonumber \\
&\times& \frac{{b\left( {T,T_{\rm c} \left( \epsilon  \right)} \right)^{2.5} }}{{k\left( {T,T_{\rm c} \left( \epsilon  \right)} \right)^2 }}h^{0.5} \left( {1 - h} \right)^2.
\end{eqnarray}}
Alternatively, one can define a strain dependent pre-constant $C(\epsilon)$ as will be shown below. The temperature dependencies in \eref{normalizedt} can be empirical (e.g. \eref{summershc2t}) or microscopic based alternatives (e.g. \eref{MDG} and \eref{kappat}). The discussion above is used in the scaling relations that are available throughout the literature.
\subsubsection{The Summers relation}
Summers~\textsl{et al.}~\cite{Summers1991}, following previous work by Hampshire~\textsl{et al.}~\cite{Hampshire1985} and Ekin~\cite{Ekin1980}, using $\epsilon=\epsilon_{\rm a}$, defined a strain dependent pre-constant:
\begin{equation} \label{ceps}
    C\left( \epsilon  \right) = C_0 s\left( \epsilon  \right)^{0.5}.
\end{equation}
The power of 0.5 is introduced in \eref{ceps} to retain consistency with the strain dependence of the bulk pinning force as postulated by Ekin \eref{ekin2}. It can be shown that the Summers relation can be rewritten in a form consistent with the previous Section as:
{\setlength\arraycolsep{1pt}
\begin{eqnarray}
    F_{\rm p} \left( {H,T,\epsilon } \right) &=& C_0 \left[ {\mu _0 H_{\rm{c2m}} \left( 0 \right)} \right]^{0.5} s\left( \epsilon  \right) \nonumber \\
    &\times& \frac{{b\left( {T,T_{\rm c} \left( \epsilon  \right)} \right)^{2.5} }}{{k\left( {T,T_{\rm c} \left( \epsilon  \right)} \right)^2 }}h^{0.5} \left( {1 - h} \right)^2,
\end{eqnarray}}
in which the empirical temperature dependence \eref{summershc2t} is used for $b(T,T_{\rm c}(\epsilon))$ and $k(T,T_{\rm c}(\epsilon))$. The Summers relation is thus, apart from the pre-constant, consistent with \eref{kramerfphte}. It has finally to be quoted that Hampshire \textsl{et al.} stated~\cite{Hampshire1985}, in relation to substantial inhomogeneity of their samples, that the validity of their calculated $\kappa_1(T)$ results is expected to break down for $T>13.5\ \rm{ K}$. It will be shown below that this, through comparison of the Summers form \eref{summershc2t} to the microscopic based form \eref{kappat} and measured results, indeed appears to be the case.
\subsubsection{Alternative approaches}
The empirical bases in the relations proposed by Ekin and Summers \textsl{et al.}, their inaccuracy, and the extensive new $J_{\rm c}(H,T,\epsilon)$ data sets for wires that have become available over the past decade have evoked a re-analysis of overall scaling behavior in Nb$_3$Sn wires. Two alternative approaches can be distinguished, focusing either on improving the strain and temperature dependent terms~\cite{tenhaken1999,Godeke1999}, or focusing on improving the basis of the strain dependence of $C/\kappa_1(T,\epsilon)^2$ in \eref{fphte}~\cite{Cheggour2002,Keys2003,Cheggour1999}.

In the first approach it is stated that, since the strain dependence through the power-law function does not account for linearity at high compressive strains, it should be replaced by a function that does. This was done through a change in $s(\epsilon)$ from the power-law form to the Ten Haken form. Secondly, it was assumed that \eref{kramerfphte}, and thereby the Summers version, is valid but that the empirical temperature dependencies as proposed by Summers \textsl{et al.} [i.e. \eref{summershc2t}] are in error and can be improved upon. The general form \eref{fpht} was chosen in~\cite{tenhaken1999,Godeke1999} under the condition that $s(\epsilon)$ from \eref{seps} is independent of temperature, \eref{straindep4K} has general validity and $(1-a_\vartriangle(H)/d_{\rm{av}})\cong1$.

The temperature dependence in \eref{fpht} is determined by the ratio $[\mu_0H_{\rm c2}(T)]^\nu/\kappa_1(T)^\gamma$. Four-dimensional parameter least squares fits using extensive data sets on a six different wires and highly linear Kramer plots indicated $n\cong1$ (as stated by Ekin), $p\cong0.5$ and $q\cong2$ (as stated by Kramer for $h>0.2$). Since the temperature dependencies of $H_{\rm c2}$ and $\kappa_1$ were fully empirical in the Summers form, both powers $\nu$ and $\gamma$ were regarded as free parameters to allow for errors in the temperature dependencies. Overall least squares fits on the data sets indicated that $\nu\cong2$ and $\gamma\cong1$ yielded the highest accuracy. The overall description \eref{fpht} with $p=0.5$, $q=2$, $\nu=2$ and $\gamma=1$ yielded fits to experimental data with a standard deviation of about 2\% over a limited magnetic field range from 5 to 13 T, but for all temperatures (4.2 K to $T_{\rm c}$) and all investigated compressive strains ($-0.7<\epsilon_{\rm a}<0$)~\cite{Godeke2002a}. It was expected that if correct dependencies for $H_{\rm c2}(T)$ and $\kappa_1(T)$ were used, that similar accurate fits would be obtained with values for $\nu$ and $\gamma$, which are consistent with Kramer's pinning theory, i.e. $\nu=2.5$ and $\gamma=2$. However, in Section \ref{Comparison}, we will re-analyze one of these earlier data-sets with the recent new insights on inhomogeneity averaging effects and the now known temperature dependencies $H_{\rm c2}(T)$ and $\kappa_1(T)$, and show that the deviating values $\nu=2$ and $\gamma=1$ are correct.

The second approach to improve upon the Summers relation emphasizes on the different dependencies of the bulk pinning force on temperature and strain in an attempt to define a better founded combination of both dependencies.  Following initial work by Kroeger \textsl{et al.}~\cite{Kroeger1980} this results in a general relation of the form~\cite{Cheggour2002,Keys2003,Cheggour1999}:
\begin{equation}
F_{\rm p} \left( {H,T,\epsilon } \right) = A\left( \epsilon  \right)\left[ {\mu _0 H_{\rm c2}^* \left( {T,\epsilon } \right)} \right]^n h^p \left( {1 - h} \right)^q,
\end{equation}
in which $n$, $p$ and $q$ are free parameters and $A(\epsilon)$ is a function of strain alone. Several versions for $A(\epsilon)$ have been proposed [also including $\kappa_1(T,\epsilon)$] in attempts to describe strain induced modifications of the flux-line to pinning center interactions and the strain dependence of $\kappa_1$.
\subsubsection{Selected general scaling relation}
In general it can be stated that much of the developments on overall scaling behavior remains empirical and it is unlikely that this empirical basis can be fully removed. As a final remark it should be mentioned that in a four-dimensional description with a large number of free parameters it often remains a somewhat subjective choice which parameters are fixed or allowed to vary, in addition to the choice of relationships that can be implemented.

The overall scaling law that is proposed in this article will now be summarized. It is assumed that the critical current density scales according to \eref{fphte} in the most general form:
\begin{equation}\label{generalshort}
    J_{\rm c} \left( {H,T,\epsilon } \right) \cong \frac{C}{{\mu _0 H}}\frac{{\left[ {\mu _0 H_{\rm c2}^* \left( {T,\epsilon } \right)} \right]^\nu  }}{{\kappa _1 \left( {T,\epsilon } \right)^\gamma  }}h^p \left( {1 - h} \right)^q,
\end{equation}
where $h=H/H_{\rm c2}^*(T,\epsilon)$, in which $H_{\rm c2}^*$ represents the inhomogeneity (bulk) averaged critical field at which the critical current density extrapolates to zero. The microscopic based temperature dependencies are selected for the upper critical field \eref{MDGshort} and the GL parameter \eref{kappat}. This leads to the general normalized form:
{\setlength\arraycolsep{1pt}
\begin{eqnarray} \label{general}
    J_{\rm c} \left( {H,T,\epsilon } \right) &\cong& \frac{C}{{\mu _0 H}}\frac{{\left[ {\mu _0 H_{\rm{c2m}}^* \left( 0 \right)} \right]^\nu  }}{{\kappa _{1\rm m} \left( 0 \right)^\gamma  }}s\left( \epsilon  \right)^{\nu  - \alpha \gamma } \nonumber \\
    &\times& \frac{{MDG\left( t \right)^\nu  }}{{k\left( t \right)^\gamma  }}h^p \left( {1 - h} \right)^q,
\end{eqnarray}}
where $t=T/T_{\rm c}^*(\epsilon)$. The term $s(\epsilon)^{\nu  - \alpha \gamma }$ arises from the strain dependence of the upper critical field through \eref{seps} and an unknown strain dependence of the zero temperature GL parameter, defined as:
\begin{equation}
    \kappa _1 \left( {0,\epsilon } \right) \equiv \kappa _{1\rm m} \left( 0 \right)s\left( \epsilon  \right)^\alpha.
\end{equation}
The form for $s(\epsilon)$ is chosen to be the improved deviatoric strain model \eref{sarno}.

The temperature and strain dependent bulk average critical field in \eref{general} is given by a combination of the microscopic temperature dependence \eref{MDGshort} and the strain dependence \eref{hc2e}:
\begin{equation}
    H_{\rm c2}^* \left( {T,\epsilon } \right) = H_{\rm{c2m}}^* \left( 0 \right)MDG\left( t \right)s\left( \epsilon  \right),
\end{equation}
and the strain dependence of the critical temperature is defined by \eref{seps}.

The constants $p$, $q$, $\nu$, $\gamma$ and $\alpha$ will have to be determined through systematic comparison to experiments, and $\varpi=3$ is used in accordance with the experimental results of Ekin. Note that this approach significantly differs from the usual separation between the dependence of the bulk pinning force on $H_{\rm c2}^*(T)$ and $H_{\rm c2}^*(\epsilon)$. Here this is not assumed \textsl{a priori} and strain dependency is simply introduced through a strain dependent upper critical field and GL parameter. This relation will be compared to measurements in Section \ref{Comparison} to determine the constants and to test its validity.
%
%
\section{\label{Comparison} Comparison to measurements}
In this Section we will compare the general $J_{\rm c}(H,T,\epsilon)$ relation \eref{general} to earlier experimental data. Experimental details, as well as the existing data-set were described elsewhere~\cite{tenhaken1999,Godeke2001,Godeke2004}. The selected data-set for the renewed analysis are $J_{\rm c}(H,T,\epsilon)$ results which were measured on an International Thermonuclear Experimental Reactor (ITER) type, Furukawa bronze processed wire. A more detailed description of this wire is also given elsewhere~\cite{Godeke2005b}. The  results of this wire are selected because of the demonstrated high reproducibility, and since it was analyzed extensively in various laboratories around the world. It should be emphasized that the results for this wire are not unique, but representative for the scaling behavior for at least five other wires from various manufacturers~\cite{Godeke1999,tenhaken1999}. $J_{\rm c}(H,T)$ results were obtained on an ITER-type Ti-6Al-4V barrel at electric field criteria of $E_{\rm c}=10^{-5}$ to $5\times10^{-4}\ \rm{ V/m}$. $J_{\rm c}(H,T,\epsilon)$ results were obtained on a Ti-6Al-4V U-shaped bending spring at $E_{\rm c}=5\times10^{-4}\ \rm{ V/m}$ and on a Ti-6Al-4V circular bending spring (`Pacman') at $E_{\rm c}=10^{-4}\ \rm{ V/m}$. The ITER barrel and U-shaped bending spring samples were heat treated together, whereas the Pacman sample was heat treated separately. All data are corrected for resistive currents that run parallel to the superconductor at any voltage criterion and the self-field generated by the wire is also corrected for. These details can be found elsewhere~\cite{Godeke2005a}.

First the field dependence will be analyzed in detail. This results in the values for $p$ and $q$ in \eref{general},  $H_{\rm c2}^*(T)$ and $F_{\rm p \max}(T)$. Then the temperature dependence will be analyzed in detail and it will be shown that the Kramer/Summers form fails to describe the temperature dependence by the use of $\nu=2.5$ and $\gamma=2$ in \eref{general}. The use of these constants results in a temperature dependent variation of $\kappa_1$ that is significantly larger than can be expected from the microscopic theory. Next the improved temperature dependence, through the use of different values for $\nu$ and $\gamma$, will be introduced and validated. This is followed by an analysis of the strain dependence using the improved deviatoric strain model. By using correct values for $\nu$ and $\gamma$, it is not required to separate the temperature and strain dependence of the bulk pinning force. Also the strain dependence of $F_{\rm p \max}$ is determined from the measurements. Finally, the minimum required data-set to determine the critical surface of an unknown wire will be discussed.
\subsection{\label{Fp/Fpmax} Normalized bulk pinning force}
\begin{figure*}
\includegraphics[scale=1]{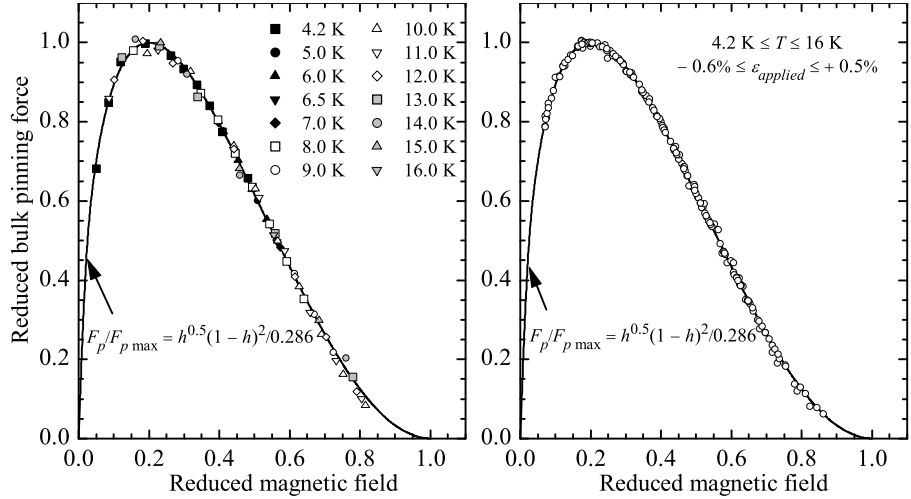}
\caption{\label{pincurve} Reduced bulk pinning force as function of reduced magnetic field for $J_{\rm c}(H,T)$ measurements at $E_{\rm c}=10^{-5}\ \rm{ V/m}$ (left graph) and $J_{\rm c}(H,T,\epsilon)$ measurements at $E_{\rm c}=5\times10^{-4}\ \rm{ V/m}$ (right graph).}
\end{figure*}

To investigate the magnetic field dependence of the bulk pinning force, each combination of temperature and strain was attributed to a unique but unknown effective upper critical field $H_{\rm c2}^*(T,\epsilon)$ and maximum bulk pinning force $F_{\rm p \max}(T,\epsilon)$. The bulk pinning forces were calculated from the measured critical current densities through $F_{\rm p}(H,T,\epsilon)=|J_{\rm c}(H,T,\epsilon)\mu_0H|$. Kramer plots are linear for this wire as will be shown below, indicating $q=2$ and $p=0.5$ as was shown in Section \ref{F-H}. To analyze consistency with these values in \eref{pinningfunction} for all temperatures and strains, a least squares fit was made to $F_{\rm p}(H,T,\epsilon)/F_{\rm p \max}(T,\epsilon)$ in:
\begin{equation}
\frac{{F_{\rm p} \left( {H,T,\epsilon } \right)}}{{F_{\rm p \max } \left( {T,\epsilon } \right)}} = \frac{{h^{0.5} \left( {1 - h} \right)^2 }}{{0.2^{0.5} \left( {1 - 0.2} \right)^2 }} \cong \frac{{h^{0.5} \left( {1 - h} \right)^2 }}{{0.286}},
\end{equation}
where $h=H/H_{\rm c2}^*(T,\epsilon)$. The resulting normalized bulk pinning force dependencies on reduced field are shown in figure \ref{pincurve}.

\begin{figure}
\includegraphics[scale=1]{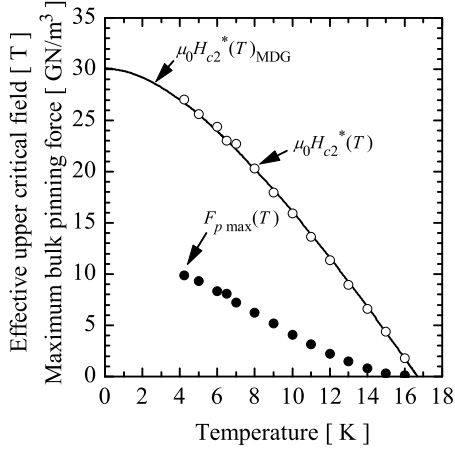}
\caption{\label{hc2fp-T} Temperature dependence of the effective upper critical field and maximum bulk pinning force as measured on the ITER barrel at $E_{\rm c}=10^{-5}\ \rm{ V/m}$.}
\end{figure}

The left graph shows the dependence for the $J_{\rm c}(H,T)$ measurements at all temperatures (and constant strain). The right graph shows the reduced dependencies for the $J_{\rm c}(H,T,\epsilon)$ for all temperature and strain values. The fit values for $H_{\rm c2}^*(T)$ and $F_{\rm p \max}(T)$ for the $J_{\rm c}(H,T)$ measurements are summarized in figure \ref{hc2fp-T}, including a calculated dependence for $H_{\rm c2}^*(T)$ using the Maki-De Gennes relation \eref{MDG} with $\mu_0H_{\rm c2}^*(0)=30.1\ \rm{ T}$ and $T_{\rm c}^*(0)=16.7\ \rm{ K}$. It should be noted that these are at $-0.15\%$ axial pre-compression, as will be shown in Section \ref{Overall}. The high value for $H_{\rm c2}^*(0)$ in comparison to a measured maximum detectable value $\mu_0H_{\rm c2}(0)=29.3\ \rm{ T}$~\cite{Godeke2005b} can be explained by inhomogeneities causing a downward tail in Kramer plots thereby resulting in an overestimate of the extrapolated $H_{\rm c2}^*(0)$. It should be emphasized that this high value is required to accurately describe the measurements in the range $h<0.8$.

The bulk pinning force curves in figure \ref{pincurve} demonstrate that $p=0.5$ and $q=2$ indeed yield the correct magnetic field dependence for the entire investigated temperature and strain regime. Above $h\cong0.8$ deviations can occur, but these are attributed to inhomogeneities. It should be emphasized that the observed correspondence with $p=0.5$ and $q=2$ is not specific for this wire, but holds for all our earlier investigated sample materials and also for `normally' reacted PIT wires as demonstrated by Cooley \textsl{et al.}~\cite{Cooley2002}. Note also that the regime above $h=0.8$ can be accounted for in the fit by using deviating values for $p$ and $q$. This will obviously lead to a different value for the scaling field $H_{\rm c2}^*$ and scaling, as demonstrated by a normalized pinning function, is still possible as was shown by similar analysis on the same wire by Hampshire and co-authors~\cite{Hampshire2001}. This will improve the accuracy of the fit above $h=0.8$ at the cost of generality of the model. However, since this regime involves very low current densities that are of negligible importance to the applications, we choose to retain a general model at the cost of low current density fit accuracy.

It is thus concluded that the field dependence is accurately described by $p=0.5$ and $q=2$ for all magnetic fields in the range $0.05<h<0.8$ for all temperatures and strains. It is also observed that the Maki-De Gennes relation indeed yields an accurate description for the temperature dependence of the effective upper critical field resulting from transport critical current measurements.
\subsection{\label{Fp-T} Temperature dependence}
The temperature dependence of the bulk pinning force can now be analyzed. The temperature dependencies for the upper critical field and the GL parameter are known. This means that only the temperature dependence of the maximum bulk pinning force has to be investigated to determine the overall temperature dependence of the bulk pinning force. The temperature dependence of the maximum bulk pinning force is determined by the values for $\nu$ and $\gamma$ in \eref{general}.
\subsubsection{Incorrect temperature dependence of the Kramer/Summers form}
In the Kramer/Summers form $\nu=2.5$ and $\gamma=2$ in \eref{general} and \eref{generalshort}, leading to:
\begin{equation}\label{sum1}
    F_{\rm p} \left( {H,T} \right) = C_1 \frac{{MDG\left( t \right)^{2.5} }}{{k\left( t \right)^2 }}h^{0.5} \left( {1 - h} \right)^2,
\end{equation}
where $C_1=C[\mu_0H_{\rm c2}^*(0)]^{2.5}/\kappa_1(0)^2$. The maximum pinning force is thus given by:
{\setlength\arraycolsep{1pt}
\begin{eqnarray}\label{sum2}
   F_{\rm p \max } \left( T \right) &\cong& 0.286C\frac{{\left[ {\mu _0 H_{\rm c2}^* \left( T \right)} \right]^{2.5} }}{{\kappa _1 \left( T \right)^2 }} \nonumber \\
 &\cong& 0.286C\frac{{\left[ {\mu _0 H_{\rm c2}^* \left( 0 \right)} \right]^{2.5} }}{{\kappa _1 \left( 0 \right)^2 }}MDG\left( t \right)^{0.5} \left( {1 - t^2 } \right)^2. \nonumber \\
 \phantom{}
\end{eqnarray}}
There is unfortunately no direct way to extract either $C$ or $\kappa_1(0)$ separately from the $J_{\rm c}(H,T)$ results. Comparison to the measurements delivers $C_1$, $H_{\rm c2}^*(0)$, $T_{\rm c}^*(0)$, $H_{\rm c2}^*(T)$ and $F_{\rm p \max}(T)$ without free parameters, but $\kappa_1(0)$ depends on the value of $C$. Note that $C$ is not a fundamental constant here because $F_{\rm p}$ is evaluated from the non-copper cross-section.

\begin{figure}
\includegraphics[scale=1]{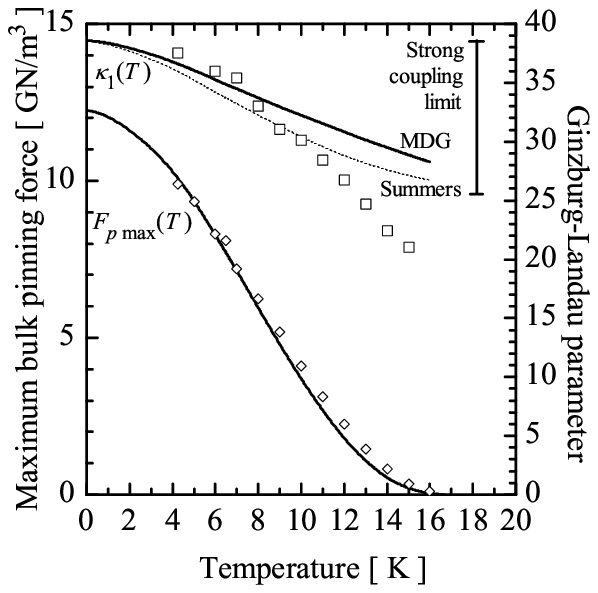}
\caption{\label{sumtdep} Temperature dependence of the GL parameter and maximum bulk pinning force at $E_{\rm c}=10^{-5}\ \rm{ V/m}$. The points are derived from the $J_{\rm c}(H,T)$ characterizations and the lines are the calculated dependencies using \eref{kappat} (MDG) and \eref{sum2}. For completeness, also the Summers empirical form for the GL parameter \eref{summershc2t} is included.}
\end{figure}

An overall least squares fit of \eref{sum1} on the measured $J_{\rm c}(H,T)$ data, using $\mu_0H_{\rm c2}^*(0)=30.1\ \rm{ T}$ and $T_{\rm c}^*(0)=16.7\ \rm{ K}$, results in $C_1=42.8 \times 10^9 \ \rm{AT/m}^2$. The resulting maximum bulk pinning force according to \eref{sum2} is shown in comparison to the actual data in figure~\ref{sumtdep}. Significant deviations occur above 9 K, which will also be present in calculated critical current densities.

The deviations in $F_{\rm p \max}(T)$ can be illustrated through the resulting change in $\kappa_1$ with temperature. The variation in $\kappa_1$ with temperature, according to the Kramer/Summers form, can be calculated through \eref{sum2} by using values for $F_{\rm p \max}(T)$ and $H_{\rm c2}^*(T)$ from figure \ref{hc2fp-T}, and by inserting a value for $\kappa_1(0)$. Choosing $C=12.8\times10^9\ \rm{ AT}^{1.5}\rm{m}^2$ as proposed by Kramer \eref{kramerpinning}, results through \eref{sum1} in $\kappa_1(0)=38.6$; a reasonable value compared to the literature~\cite{Guritanu2004}. The resulting values for $\kappa_1(T)$, calculated from $F_{\rm p \max}(T)$ and $H_{\rm c2}^*(T)$ using \eref{sum2} are plotted in figure \ref{sumtdep}, together with the microscopic \eref{kappat} and empirical \eref{summershc2t} dependencies.

The change of $\kappa_1(T)$ of nearly 100\% is unrealistic, since a maximum change of 20\% is expected in the weak, and 50\% in the strong coupling limit as discussed in Section \ref{Hc2-T}. This is independent of the choice of $C$ and thus $\kappa_1(0)$, since the calculations are normalized to $\kappa_1(0)$. The combination of $\nu=2.5$ and $\gamma=2$ therefore results in an unrealistically large change in $\kappa_1(T)$, calculated from $F_{\rm p \max}(T)$ and $H_{\rm c2}^*(T)$.

\begin{figure}
\includegraphics[scale=1]{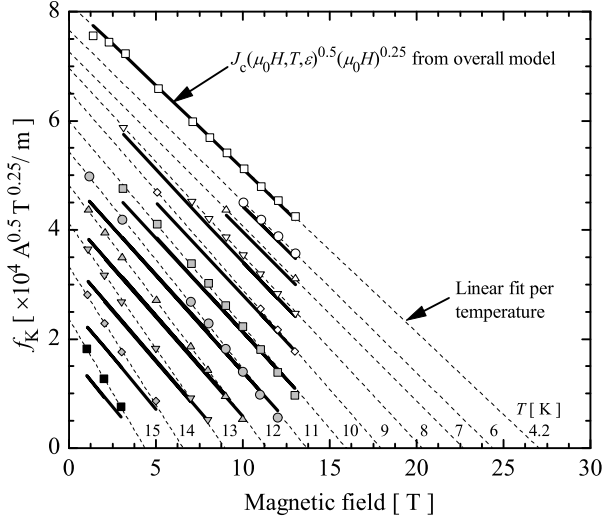}
\caption{\label{sumkram} Kramer plot of measured $J_{\rm c}(H,T)$ data, including linear fits according to \eref{kramerplot} and the overall $J_{\rm c}(H,T)$ fit according to \eref{sum1} at $E_{\rm c}=10^{-5}\ \rm{ V/m}$. The graph emphasizes the failure of the Kramer/Summers model to account for the correct slope at higher temperatures, which leads to significant deviations in the calculated $J_{\rm c}(H,T)$ values.}
\end{figure}

The Kramer function is given by \eref{kramerplot} and the slope is therefore $C/\kappa_1(T)$. The large change in $\kappa_1$ with temperature thus translates to a larger change in the slope of a Kramer plot at higher temperatures. This is visualized by the Kramer plot given in figure \ref{sumkram}.

The overall $J_{\rm c}(H,T)$ least squares fit is optimized at 4.2 K in this graph, to emphasize the deviations that occur at higher temperatures. The error in the overall temperature dependence in the Kramer/Summers model, through the use of $\nu=2.5$ and $\gamma=2$, causes significant deviations in the slope of a Kramer plot at higher temperatures and thus in large errors in the calculated $J_{\rm c}(H,T)$ values.

Deviations in the Kramer fit were observed also in our earlier analysis~\cite{Godeke1998,Godeke2000} and were attributed to incorrect (empirical) temperature dependencies of the effective upper critical field and the GL parameter. Now that $H_{\rm c2}^*(T)$ and $\kappa_1(T)$ are replaced by microscopic based alternatives, it has to be concluded that the observable errors in $F_{\rm p \max}(T)$ stem from incorrect values for $\nu$ and $\gamma$. This is supported by the unrealistic large change in the calculated $\kappa_1$ with temperature, surpassing the strong coupling limit.
\subsubsection{Improved temperature dependence}
\begin{figure}
\includegraphics[scale=1]{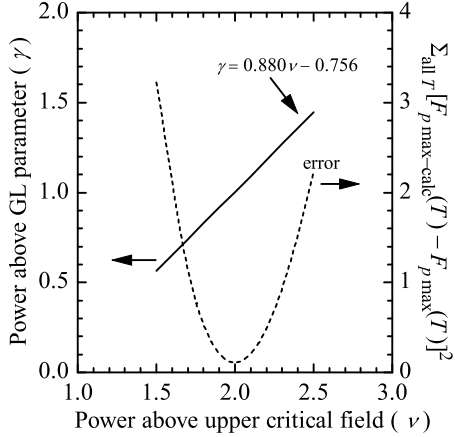}
\caption{\label{nugamma} Combinations of $\nu$ and $\gamma$ and the overall least squares fit error in the maximum bulk pinning force.}
\end{figure}

\begin{figure}
\includegraphics[scale=1]{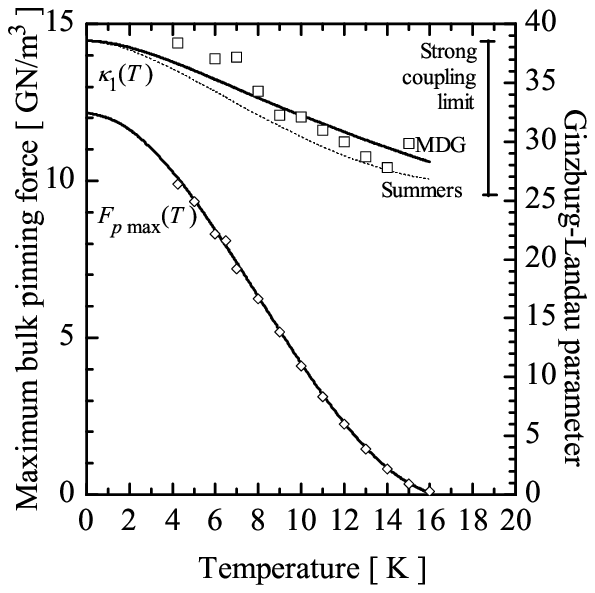}
\caption{\label{Arnotdep} Improved temperature dependence of the GL parameter and maximum bulk pinning force at $E_{\rm c}=10^{-5}\ \rm{ V/m}$. The points are derived from $J_{\rm c}(H,T)$ characterizations and the lines are the calculated dependencies using \eref{kappat} (MDG) and \eref{generalfpt} with $\nu=2$ and $\gamma=1$. For completeness, also the Summers empirical form for the GL parameter \eref{summershc2t} is included.}
\end{figure}

In the above analysis it became clear that the temperature dependence of the bulk pinning force using $\nu=2.5$ and $\gamma=2$ is incorrect. It is therefore required to determine the optimal values for both constants that yield the highest accuracy for the calculation of the temperature dependence of the maximum bulk pinning force. The $F_{\rm p \max}(T)$ data are, in the general form, described by:
{\setlength\arraycolsep{1pt}
\begin{eqnarray}\label{generalfpt}
   F_{\rm p \max } \left( T \right) &\cong& 0.286C\frac{{\left[ {\mu _0 H_{\rm c2}^* \left( T \right)} \right]^\nu  }}{{\kappa _1 \left( T \right)^\gamma  }} \nonumber \\
  &\cong& 0.286C\frac{{\left[ {\mu _0 H_{\rm c2}^* \left( 0 \right)} \right]^\nu  }}{{\kappa _1 \left( 0 \right)^\gamma  }}MDG\left( t \right)^{\nu  - \gamma } \left( {1 - t^2 } \right)^\gamma. \nonumber \\
  \phantom{}
\end{eqnarray}}
Relation \eref{generalfpt} is fitted against the $F_{\rm p \max}(T)$ data from Section \ref{Fp/Fpmax}. Values of $\nu$ are chosen around 2.5 and the value for $\gamma$ is least squares fitted. Various combinations of $\nu$ and $\gamma$ are able to describe the measured $F_{\rm p \max}(T)$, each with a specific remaining overall least squares error, as plotted in figure \ref{nugamma}. The possible combinations of $\nu$ and $\gamma$ are related by $\gamma=0.880\nu-0.756$. The overall error shows a distinct minimum of nearly zero for the combination $\nu=2$ and $\gamma=1$ and these values therefore yield optimum accuracy for the temperature dependence of the maximum bulk pining force. Relation \eref{generalfpt}, using $\nu=2$ and $\gamma=1$ is plotted together with the $F_{\rm p \max}(T)$ data from Section \ref{Fp/Fpmax} in figure \ref{Arnotdep}. The accuracy in the calculated $F_{\rm p \max}(T)$ is significantly improved compared to the use of $\nu=2.5$ and $\gamma=2$ in figure \ref{sumtdep}.

\begin{figure}
\includegraphics[scale=1]{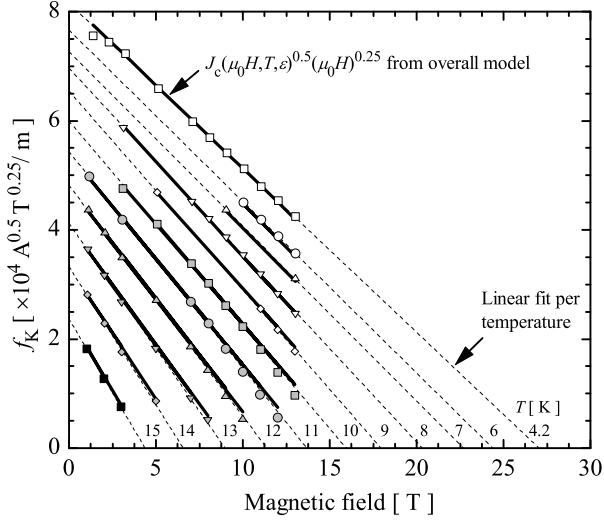}
\caption{\label{arnokram} Kramer plot of measured $J_{\rm c}(H,T)$ data at $E_{\rm c}=10^{-5}\ \rm{ V/m}$, including linear fits according to \eref{7-7} and the overall $J_{\rm c}(H,T)$ fit according to \eref{7-6}. The improved overall temperature dependence using $\nu=2$ and $\gamma=1$ results in an accurate description across the entire temperature range.}
\end{figure}

The temperature dependence of the GL parameter, resulting from $F_{\rm p \max}(T)$ and $H_{\rm c2}^*(T)$ using \eref{generalfpt} with $\nu=2$ and $\gamma=1$ can also be calculated. The relation between $C$ and $\kappa_1(0)$ now differs from the Kramer form \eref{kramerpinning}. The value for $\kappa_1(0)$ is chosen consistent with the previous analysis, i.e. $\kappa_1(0)=38.6$, leading to $C=1.81\times10^9\ \rm{ AT}^{1.5}\rm{m}^2$. This calculation yields a $\kappa_1(T)$ as plotted in figure \ref{Arnotdep}. The GL parameter now changes by about 35\% which is in agreement with expectations (i.e. between 20 and 50\%, see Section \ref{microkappa}) and, moreover, follows the microscopic based description.
The overall accuracy of the temperature dependence can again also be demonstrated using a Kramer plot. Using $\nu=2$ and $\gamma=1$ in \eref{general} results in an overall description:
{\setlength\arraycolsep{1pt}
\begin{eqnarray}\label{7-6}
   F_{\rm p} \left( {H,T} \right) &=& \frac{{C\left[ {\mu _0 H_{\rm c2}^* \left( 0 \right)} \right]^2 }}{{\kappa _1 \left( 0 \right)}}\frac{{MDG\left( t \right)^2 }}{{k\left( t \right)}}h^{0.5} \left( {1 - h} \right)^2  \nonumber \\
   &=& C_1 MDG\left( t \right)\left( {1 - t^2 } \right)h^{0.5} \left( {1 - h} \right)^2,
\end{eqnarray}}
and the linear Kramer function becomes:
\begin{equation}\label{7-7}
    f_{\rm K} \left( {H,T} \right)  = \frac{{C^{0.5} }}{{\kappa _1 \left( T \right)^{0.5} \left[ {\mu _0 H_{\rm c2}^* \left( T \right)} \right]^{0.25} }}\mu _0 \left( {H_{\rm c2}^* \left( T \right) - H} \right).
\end{equation}
Similar analysis as in the previous Section yields from the overall $J_{\rm c}(H,T)$ least squares fit $C_1=42.6\times10^9\ \rm{ AT/m}^2$. The small difference in $C_1$ compared to the Kramer/Summers form ($42.8\times10^9\ \rm{ AT/m}^2$) stems from the fact that in figure \ref{sumkram} the fit was optimized for the 4.2 K results, whereas now the fit is optimized over the entire temperature range. The improved functionality yields slopes that are consistent with the measured results over the entire temperature range as is seen in figure \ref{arnokram}. The only significant deviations occur close to $H_{\rm c2}^*(T)$, the regime where non-linearities in Kramer plots originate from A15 inhomogeneities.
\subsection{\label{Fp-eps} Strain dependence}
Now that the magnetic field and temperature dependence are established, the strain dependence is introduced by stating that the effective upper critical field, the critical temperature and the GL parameter depend on strain. This leads through \eref{generalshort} to:
\begin{equation}\label{7-8}
    F_{\rm p} \left( {H,T,\epsilon } \right) = C\frac{{\left[ {\mu _0 H_{\rm c2}^* \left( {T,\epsilon } \right)} \right]^2 }}{{\kappa _1 \left( {T,\epsilon } \right)}}h^{0.5} \left( {1 - h} \right)^2,
\end{equation}
where  $H_{\rm c2}^*(T,\epsilon)\equiv H_{\rm{c2m}}^*(T)s(\epsilon)$, $T_{\rm c}^*(\epsilon)\equiv T_{\rm{cm}}^*s(\epsilon)^{1/\varpi}$ and $\kappa_1(T,\epsilon)\equiv \kappa _{1\rm m}(T)s(\epsilon)^\alpha$. It is assumed that $\varpi=3$ since this was shown earlier to be consistent with measurements, and $\alpha$ has to be determined from experiment. In normalized form \eref{7-8} becomes [through \eref{general}]:
\begin{equation}\label{7-9}
    F_{\rm p} \left( {H,T,\epsilon } \right) = C_1 s\left( \epsilon  \right)^{2 - \alpha } MDG\left( t \right)\left( {1 - t^2 } \right)h^{0.5} \left( {1 - h} \right)^2,
\end{equation}
where $C_1=C[\mu_0H_{\rm{c2m}}^*(0)]^2/\kappa _{1\rm m}(0)$.

\begin{figure}
\includegraphics[scale=1]{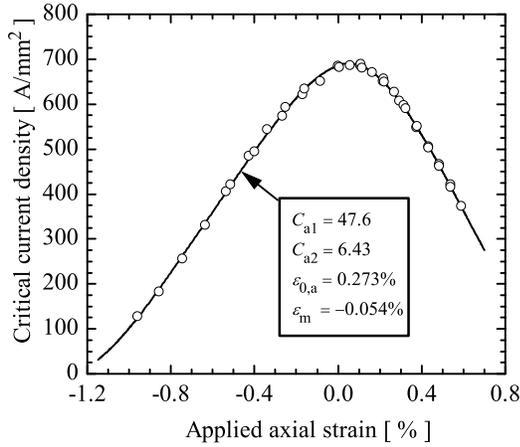}
\caption{\label{pacman} Axial strain dependence of the critical current density at $E_{\rm c}=10^{-4}\ \rm{ V/m}$, 12 T and 4.2 K measured on a circular bending spring and calculated from \eref{7-9} and \eref{sarno} using parameters as depicted in the graph.}
\end{figure}

First the function for $s(\epsilon)$ will be determined. It will initially be assumed that $\alpha=1$, since this is consistent with Ekin's~\cite{Ekin1980} statement that $F_{\mathrm p \max}(4.2~\mathrm{K})\propto H_{\mathrm c2}^*(\epsilon)$, as well as our earlier analysis. This choice for $\alpha$ has to be confirmed with experimental results. The improved and re-normalized deviatoric strain model is chosen as description for $s(\epsilon)$, i.e. relation \eref{sarno}. An accurate $J_{\rm c}(H,T,\epsilon)$ measurement with a high strain resolution is required to determine the parameters in \eref{sarno}, i.e. $C_{\rm a1}$, $C_{\rm a2}$, $\epsilon_{0,\rm a}$ and $\epsilon_{\rm m}$. The mechanical parameters are therefore determined from a measurement on the Furukawa wire at 12 T and 4.2 K, which are measured with the required strain resolution~\cite{Godeke2004}, as shown in figure \ref{pacman}. The points are the $J_{\rm c}(\epsilon)$ results at 12 T and 4.2 K at a criterion of $E_{\rm c}=10^{-4}\ \rm{ V/m}$. The line is calculated from \eref{7-9} combined with \eref{sarno} using the mechanical parameters depicted in the graph, $\alpha=1$, $C_1=47.8\ \rm{ kAT/mm}^2$, $\mu_0H_{\rm{c2m}}^*(0)=30.7\ \rm{ T}$ and $T_{\rm{cm}}^*(0)=16.8\ \rm{ K}$.

It should be emphasized that $C_1$, $H_{\rm{c2m}}^*(0)$ and $T_{\rm{cm}}^*(0)$ are solely used as fit parameters to scale $J_{\rm c}$. Since no field or temperature dependent measurements were performed on this specific sample they are arbitrary, but not unique, parameters resulting from a least squares fit of the 12 T, 4.2 K data points. As mentioned this sample was not heat treated together with the U-spring and barrel samples, rendering differences of a few percent in $C_1$,  $H_{\rm{c2m}}^*(0)$ and $T_{\rm{cm}}^*(0)$ possible. In addition, the applied criterion of $E_{\rm c}=10^{-4}\ \rm{ V/m}$ is higher than used for the $J_{\rm c}(H,T)$ characterizations ($E_{\rm c}=10^{-5}\ \rm{ V/m}$), translating to slightly higher values for the fit parameters, since effectively Sn richer A15 sections will be probed.

Figure \ref{pacman} shows that the improved deviatoric strain model does accurately account for the measured behavior. In contrast to the original deviatoric strain model, it is able to fit the entire strain curve and not just the compressive part. Note that the term $C_{\rm a2}$ effectively tilts the calculated curve around $H_{\rm{c2m}}^*(T)$ at $\epsilon_{\rm a}=0$. This causes the fit values for $C_{\rm a1}$ and $\epsilon_{0,\rm a}$ to change compared to the earlier model. Overall, however, \eref{sarno} accurately describes the strain dependence and $\alpha=1$ in \eref{7-9} appears to yield the correct strain dependence of the bulk pinning force.

\begin{figure}
\includegraphics[scale=1]{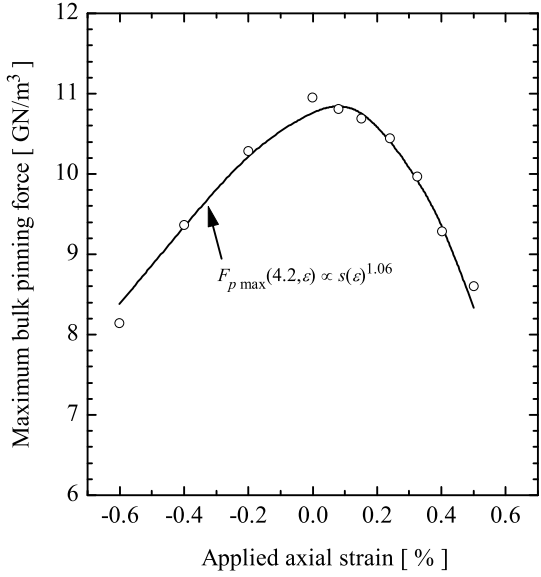}
\caption{\label{maxpinstrain} Maximum bulk pinning force as function of applied axial strain at $E_{\rm c}=5\times10^{-4}\ \rm{ V/m}$ and 4.2 K, measured on the U-spring.}
\end{figure}

Now that $s(\epsilon)$ is confirmed to be accurate, $F_{\rm p \max}(\epsilon)\propto s(\epsilon)^{2-\alpha}$ has to be investigated to find the value for $\alpha$ from experiment, which can be done using earlier U-spring results. Note that experimental limitations of the U-spring only allow for a relatively high criterion of $E_{\rm c}=5 \times 10^{-4}\ \rm{ V/m}$ to be considered reliably~\cite{Godeke2004}. The power $2-\alpha$ is determined by plotting the $F_{\rm p \max}(\epsilon)$ values from Section \ref{Fp/Fpmax} versus axial strain and comparing to $s(\epsilon)^{2-\alpha}$. This is demonstrated in figure \ref{maxpinstrain} for 4.2 K results, yielding $\alpha=1.0\pm0.1$, in agreement with Ekin's result. Similar analyses for higher temperatures suggest that $\alpha$ might decrease with temperature. At higher temperatures, however, the $F_{\rm p \max}(\epsilon)$ analyses become significantly less accurate since they are based on much smaller current densities and limited data. Also inhomogeneity effects, which manifest themselves in the investigated magnetic field range at higher temperatures, complicate a more precise determination of $2-\alpha$. The results are therefore not sufficiently accurate to come to unambiguous conclusions on a possible temperature dependence of $\alpha$. Moreover, strain dependency results are, for now, only available for inhomogeneous samples, rendering further speculations not very useful.  With the available experimental results it can be assumed that $\alpha\cong1$ for all temperatures.

It is concluded that strain dependence can be introduced in the bulk pinning force using a single dependence $F_{\rm p \max}\propto H_{\rm c2}^*(T,\epsilon)^2$, provided that $\alpha\cong1$ for all temperatures. This is in contrast to the usual conclusion that $H_{\rm c2}^*(T)$ and $H_{\rm c2}^*(\epsilon)$ have different dependencies. This single function is enabled through the use of values for $\nu$ and $\gamma$ that accurately agree with measured results and by including the strain dependence of the GL parameter. Figure \ref{maxpinstrain} implies that $\kappa_1$ has a similar strain dependence as $H_{\rm c2}^*$.
%
%
\section{\label{Overall} Scaling of measured results}
In the previous Sections an overall scaling function for $F_{\rm p}(H,T,\epsilon)$ was postulated and confirmed by systematic analysis of the separate dependencies on magnetic field, temperature and strain. The proposed function is given by \eref{7-9}, with $\alpha\cong1$. Solving the implicit Maki-De Gennes relation can be avoided by using its approximation $MDG(t)\cong(1-t^{1.52})$ which is sufficiently accurate for practical applications. Using this approximation changes the overall function to the simple form:
{\setlength\arraycolsep{1pt}
\begin{eqnarray}\label{7-10}
   &J_{\rm c} \left( {H,T,\epsilon } \right) \cong \frac{{C_1 }}{{\mu _0 H}}s\left( \epsilon  \right)\left( {1 - t^{1.52} } \right)\left( {1 - t^2 } \right)h^{0.5} \left( {1 - h} \right)^2,   \nonumber \\
   &\rm{with} \nonumber \\
   &C_1  = {{C\left[ {\mu _0 H_{\rm{c2m}}^* \left( 0 \right)} \right]^2 } \mathord{\left/ {\vphantom {{C\left[ {\mu _0 H_{\rm{c2m}}^* \left( 0 \right)} \right]^2 } {\kappa _{1\rm m} \left( 0 \right)}}} \right. \kern-\nulldelimiterspace} {\kappa _{1\rm m} \left( 0 \right)}}, \nonumber \\
   &t \equiv {T \mathord{\left/ {\vphantom {T {T_{\rm c}^* \left( \epsilon  \right)}}} \right.\kern-\nulldelimiterspace} {T_{\rm c}^* \left( \epsilon  \right)}},{\qquad}h \equiv H/H_{\rm c2}^* \left( {T,\epsilon } \right), \nonumber \\
   &H_{\rm c2}^* \left( {T,\epsilon } \right) \cong H_{\rm{c2m}}^* \left( 0 \right)s\left( \epsilon  \right)\left( {1 - t^{1.52} } \right), \nonumber \\
   &T_{\rm c}^* \left( \epsilon  \right) = T_{\rm{cm}}^* s\left( \epsilon  \right)^{\frac{1}{3}},
\end{eqnarray}}
in which $s(\epsilon)$ is given by \eref{sarno}. The superconducting parameters $C_1$, $H_{\rm{c2m}}^*(0)$ and $T_{\rm{cm}}^*(0)$ and the deformation related parameters $C_{\rm a1}$, $C_{\rm a2}$, $\epsilon_{0,\rm a}$ and $\epsilon_{\rm m}$ have to be determined experimentally.

The new scaling relation can be applied to the entire existing Furukawa wire $J_{\rm c}(H,T,\epsilon)$ data-set at $E_{\rm c}=5\times10^{-4}\ \rm{ V/m}$ which results in the mechanical and superconducting parameters given in table \ref{fitparameters}. The axial thermal pre-compression ($\epsilon_{\rm m}$) of the A15 volumes in the Furukawa wire when mounted on the specific experimental setups are given in table \ref{precompression}. These axial pre-compression values are low compared to a wire that is not mounted on a substrate (for which the pre-compression usually is between $-0.2$ and $-0.4$\%). This is a result of the low thermal contraction of the substrate material (Ti-6Al-4V) which determines the overall thermal contraction after wire mounting and which matches the thermal contraction of Nb$_3$Sn. Differences between the thermal pre-compression on the ITER barrel and the strain devices are attributed to the differences in mounting temperature of the wire (epoxy hardening at room temperature for the ITER barrel versus soldering at about 200$^\circ$C for the strain devices).
\begin{table}
    \caption{\label{fitparameters} Parameters for the calculation of $J_{\rm c}(H,T,\epsilon)$ at $E_{\rm c}=5\times10^{-4}\ \rm{ V/m}$.}
    \begin{indented}   
        \item[]    
        \begin{tabular}{cccccc}
            \br    
            \centre{3}{Deformation}&\centre{3}{Superconducting} \\ 
            \crule{3}&\crule{3} \\ 
            $C_{\rm a1}$ & $C_{\rm a2}$ & $\epsilon_{0,\rm a}$ & $\mu_0H_{\rm{c2m}}^*(0)$ & $T_{\rm{cm}}^*(0)$ & $C_1$ \\
            \mr    
            47.6 & 6.4 & 0.273 & 30.7 & 16.8 & 46.3 \\
            \br    
        \end{tabular}
    \end{indented} 
\end{table}
\begin{table}
    \caption{\label{precompression} Axial thermal pre-compression in percent at $T=4.2 \ \rm{ K}$ for a Furukawa wire mounted on an experimental setup.}
    \begin{indented}   
        \item[]    
        \begin{tabular}{ccc}
            \br    
            {ITER barrel} & {U-spring} & {Pacman}  \\
            \mr    
            $-0.153$ & $-0.045$ & $-0.054$ \\
            \br    
        \end{tabular}
    \end{indented} 
\end{table}
%
%
\section{\label{Discussion} Discussion}
\subsection{\label{Accuracy} Scaling accuracy}
Through systematic comparison with measured results deficiencies in existing scaling relations can be highlighted. Improvements can, however, only be made empirically through a change of the bulk pinning force dependency on both $H_{\rm c2}^*(T)$ and $\kappa_1(T)$ simultaneously. The resulting relation describes the measurements accurately (also for a multitude of other wires)~\cite{tenhaken1999,Godeke1999}. The values $\nu=2$ and $\gamma=1$ are similar to those found earlier~\cite{tenhaken1999,Godeke1999,Godeke2001} through minimization of the error in the overall $J_{\rm c}(H,T,\epsilon)$ least squares fit. In the earlier analysis it was assumed that the deviating values for $\nu$ and $\gamma$ originated from a wrong temperature dependence of $H_{\rm c2}$ and $\kappa_1$. It is now concluded that the Kramer form, which results from the Labusch~\cite{Labusch1969b} function for $C_{66}$, is in fact not accurate. These theories have been questioned in the literature, specifically by Brandt~\cite{Brandt1977} and Dew-Hughes~\cite{Dew-Hughes1974}, emphasizing the difficulties in understanding and modeling the flux-line to pinning site interactions. A reasonable temperature dependence description can be achieved for the combinations $\gamma=0.880\nu-0.756$, but the overall error shows a distinct minimum at $\nu=2$ and $\gamma=1$. For a proper connection to theory these values should result from a fundamental form of these interactions. At present, to our knowledge, such a model is not available.

The validity of the new scaling relation was tested on a database derived from four samples, one of which was heat treated separately. The samples were characterized in three experimental setups. Deriving a singular parameter set for the Furukawa wire is complicated by small differences between the samples and characterization methods. A better fit to the U-spring strain results alone can, for example, be found by the use of slightly different parameters. This stems from small, for now unclear, differences between the separate samples and experiments. Ideally all measurements should be performed on a single instrument and on a single wire and at one single, low criterion. The latter is preferred to remove inaccuracies arising from parasitic current subtraction and for consistency with the more standard criterion of $E_{\rm c}=10^{-5}\ \rm{ V/m}$. Nevertheless, the accuracy of a very similar function (using the earlier deviatoric strain model) was investigated extensively in previous work and it was concluded that a standard deviation below 2\% can be obtained~\cite{Godeke2002a}. The applicability of the proposed scaling function was validated for Internal-Tin wires during the ITER benchmark tests in our earlier work~\cite{Godeke1999}, and in unpublished analysis of Powder-in-Tube wires.
\subsection{\label{Minimum} Minimum required dataset}
With a scaling relation available the question arises as to what is the minimum data-set that is needed to retrieve the entire critical surface. For the temperature dependence two $H_{\rm c2}^*(T)$ points are required to determine the entire $H_{\rm c2}^*(T)$ dependence through the MDG relation. These should be determined by transport $J_{\rm c}$ measurements since these yield the correct $H_{\rm c2}^*$ and $T_{\rm c}^*$ required for scaling. From figure \ref{arnokram} it is seen that preferably at both ends of the temperature range the slope of the Kramer plot has to be determined to extrapolate to $H_{\rm c2}^*(T)$. This is most simply done by a $J_{\rm c}(H)$ measurement at 4.2 K at a few magnetic field values to check for linearity and thus validate the choice of $q=2$. In addition this should be done at a higher temperature, e.g. 12 K might be a good compromise between temperature range and available magnetic field in a laboratory magnet. Non-linearities above $h\cong0.8$ should be neglected since they most probably result from inhomogeneities. The results can then be extrapolated linearly in a Kramer plot to yield $H_{\rm c2}^*(4.2\ \rm{ K})$ and $H_{\rm c2}^*(12\ \rm{ K})$. The MDG relation (or its approximation) then yields $H_{\rm c2}^*(0,\epsilon=\epsilon_{\rm m})$ and $T_{\rm c}^*(0,\epsilon=\epsilon_{\rm m})$. $J_{\rm c}(H,T,\epsilon=\epsilon_{\rm m})$ and $C_1$ are known through \eref{7-10} using a least squares fit to the $J_{\rm c}$ data-set.

The strain dependence can then be determined by a 4.2 K strain measurement with sufficient strain resolution to accurately determine the mechanical parameters (e.g. as in figure \ref{pacman}). The magnetic field for the strain characterization should preferably be chosen such, that over the entire strain range of interest, the critical current remains above about 10 A to avoid inaccuracies at low current measurements. A comparison to the $J_{\rm c}(H,4.2\ \rm{ K})$ results above will yield $\epsilon_{\rm m}$ and $H_{\rm{c2m}}^*(0)$ and $T_{\rm{cm}}^*(0)$ can be calculated. The entire critical surface up to the magnetic field where inhomogeneity effects start to appear (i.e. $h\cong0.8$) is then known. The characterizations should preferably be done using one sample and one instrument to avoid experimental errors. The required data-set can thus be summarized as follows:
\begin{enumerate}
\item An $I_{\rm c}(H)$ measurement at 4.2 K from about 5 T up to the maximum available magnetic field (see figure \ref{arnokram}), preferably 12 to 15 T.
\item An $I_{\rm c}(H)$ measurement at about 12 K from 1 T or as low as possible, up to the maximum available magnetic field. The selectable temperature depends on the available magnetic field range (see figure \ref{arnokram}).
\item An $I_{\rm c}(\epsilon)$ measurement at 4.2 K and at a magnetic field where the critical current remains significant (e.g. $>10\ \rm{ A}$) over the entire strain range of interest (see figure \ref{pacman}).
\end{enumerate}
%
%
\section{\label{Conclusions} Conclusions}
A new, generally valid scaling relation is proposed which is based on a Kramer type pinning description. A number of improvements are made in comparison to the Summers form. A first improvement is a replacement of the empirical temperature dependency for the upper critical field with a microscopic based alternative. A second improvement is to replace the empirical temperature dependence of the Ginzburg-Landau parameter by a microscopic based alternative. A third improvement is made by replacing the power-law strain dependency with a new empirical deviatoric strain description that accounts for the effects of the third strain invariant and thus for asymmetry in the axial stain dependency curve. The use of an explicit normalized strain dependence term $s({\epsilon})$ enables straightforward implementation of optional future, more fundamental based, strain descriptions in the general scaling relation. A fourth improvement involves accounting for the inhomogeneities that are present in the A15 regions in wires by recognizing that the critical current scales with a bulk, inhomogeneity averaged critical field, representing some average of the upper critical field distribution that is always present in the A15 in wires.

A systematic comparison of measured critical current as function of magnetic field, temperature and strain with scaling relations shows that flux-line shear can be assumed to be the main de-pinning mechanism and that tails in Kramer plots can reasonably be attributed to A15 inhomogeneities. This results in a generally valid magnetic field dependence of the Kramer form, i.e. $F_{\rm p} \propto h^{0.5}(1-h)^2$ within the range $0.05<h<0.8$.

Through the analysis of the magnetic field and temperature dependence of the bulk pinning force it is shown that the Kramer/Summers form, i.e. $F_{\rm p} \propto [H_{\rm c2}^*(T)]^{2.5}/\kappa_1(T)^2$, fails to describe the observed temperature dependence with sufficient accuracy. The earlier suggestion that this followed from an empirical temperature dependence of the upper critical field and the GL constant was contradicted through the use of microscopic based relations for both. The change in $\kappa_1(T)$ in the Kramer/Summers form is significantly larger than 50\% as is expected from microscopic theory in the strong coupling limit. The improved temperature dependence of the form $F_{\rm p} \propto [H_{\rm c2}^*(T)]^2/\kappa_1(T)$ does accurately describe the measurements. The resulting change in $\kappa_1(T)$ follows the microscopic form and is within the strong coupling limit. The new dependency is in disagreement with existing relations for the flux-line shear modulus $C_{66}$. It is suggested that the temperature dependence of $C_{66}$ as given by Labusch needs to be reevaluated.

The new temperature dependence of the bulk pinning force allows for an unambiguous introduction of the strain dependence of the bulk pinning force, without the need to include separate functions to describe the strain dependent upper critical field and temperature dependent upper critical field, as is usually done in the literature. The measurements indicate that the Ginzburg-Landau parameter has similar strain dependence as the upper critical field.

The improved deviatoric strain model, which includes asymmetry, yields an accurate description of measured behavior. Though empirical, it is based on conclusions resulting from more fundamental approaches and it recognizes the three dimensional nature of strain.
\section*{Acknowledgments}
This work was supported by the Netherlands Organization for Scientific Research (NWO) through the Technology Foundation STW and by the Director, Office of Science, Office of Fusion Energy Sciences, of the U.S. Department of Energy under Contract DE-AC02-05CH11231. The Furukawa wires used in this research were donated by the Japan Atomic Energy Agency (JAEA) and were characterized under contract for the ITER European Home Team. All contributions are greatly acknowledged.
%
%
\section*{References}
\end{document}